\documentclass[amssymb,amsmath,prd,twocolumn,superscriptaddress,nofootinbib]{revtex4-1}

\usepackage{graphicx}
\usepackage{bm}
\usepackage{amssymb,amsmath}
\usepackage{mathrsfs}
\usepackage{latexsym}
\usepackage{color}
\usepackage[normalem]{ulem}
\usepackage{dcolumn}
\usepackage[colorlinks=true,citecolor=blue,urlcolor=blue]{hyperref}
\usepackage[usenames,dvipsnames]{xcolor}
\usepackage{xspace}
\usepackage{caption}
\usepackage{subcaption}
\usepackage{graphicx}

\allowdisplaybreaks

\newcommand{\CCA}{\affiliation{Center for Computational Astrophysics, Flatiron Institute, 162 5th Ave, New York, NY 10010, USA}}

\newcommand{\MSFC}{\affiliation{NASA Marshall Space Flight Center, Huntsville, AL 35812, USA}}

\newcommand{\XGI}{\affiliation{eXtreme Gravity Institute, Department of Physics, Montana State University, Bozeman, Montana 59717, USA}}

\newcommand{\MelbourneOzGrav}{\affiliation{OzGrav, University of Melbourne, Parkville, Victoria 3010, Australia}}
\newcommand{\GT}{\affiliation{Center for Relativistic Astrophysics and School of Physics, Georgia Institute of Technology, Atlanta, GA 30332, USA}}

\definecolor{kcmagenta}{rgb}{0.54, 0.17, 0.88}
\definecolor{chorange}{rgb}{0.851, 0.372, 0.007}
\definecolor{tlteal}{rgb}{0,.55,.55}
\definecolor{jcpink}{rgb}{1.0, 0.0, 0.5}
\definecolor{mmgreen}{rgb}{0.0, 0.8, 0.6}
\definecolor{bbsalmon}{rgb}{1.0, 0.47, 0.42}

\newcommand{\BayesLine}{{\tt BayesLine}\xspace}
\newcommand{\BayesWave}{{\tt BayesWave}\xspace}
\newcommand{\BayesWavePost}{{\tt BayesWavePost}\xspace}
\newcommand{\bayeswavepipe}{{\tt BayesWavePipe}\xspace}

\newcommand{\GlitchBuster}{{\tt GlitchBuster}\xspace}

\newcommand{\SNR}{\rho}

\newcommand{\wavelet}{\Psi}
\newcommand{\hplus}{h_+}
\newcommand{\hcross}{h_\times}

\begin{document}

\title{The \BayesWave analysis pipeline in the era of gravitational wave observations}

\author{Neil J. Cornish} \XGI 
\author{Tyson B. Littenberg}\MSFC
%\author{ORDER APHABETICAL}
\author{Bence B\'ecsy} \XGI
\author{Katerina Chatziioannou} \CCA
\author{James A. Clark} \GT 
\author{Sudarshan Ghonge} \GT
\author{Margaret Millhouse} \MelbourneOzGrav
%\author{OTHERS}

\date{\today}

\begin{abstract}

We describe updates and improvements to the \BayesWave gravitational wave transient analysis pipeline, and provide examples of how the algorithm is used to analyze data from ground-based gravitational wave detectors. \BayesWave models gravitational wave signals in a morphology-independent manner
through a sum of frame functions, such as Morlet-Gabor wavelets or chirplets.  \BayesWave models the instrument noise using a combination of a parametrized Gaussian noise component and non-stationary and non-Gaussian noise transients. Both the signal model and noise model employ trans-dimensional sampling, with the complexity of the model adapting to the  requirements of the data. The flexibility of the algorithm makes it suitable for a variety of analyses, including reconstructing generic unmodeled signals; cross checks against modeled analyses for compact binaries; as well as separating coherent signals from incoherent instrumental noise transients (glitches). The \BayesWave model has been extended
to account for gravitational wave signals with generic polarization content and the simultaneous presence of signals and glitches in the
data. We describe updates in the \BayesWave prior distributions, sampling proposals, and burn-in stage that provide significantly improved sampling efficiency.  We present standard review checks indicating the robustness and convergence of the \BayesWave trans-dimensional sampler.

\end{abstract}

\maketitle

%%%%%%%%%%%%%%%%%%%%%%%%%%%%%%%%%%%%%%%%%%%%%%%%%%
\section{Introduction}
\label{sec:intro}
%%%%%%%%%%%%%%%%%%%%%%%%%%%%%%%%%%%%%%%%%%%%%%%%%%

The era of gravitational wave observations began in earnest in September 2015 with
the first detection of gravitational waves from a binary black hole merger~\cite{Abbott:2016blz}.
In anticipation of the detections, a new approach to gravitational wave data analysis was proposed~\cite{Cornish:2014kda,Littenberg:2014oda}
that uses trans-dimensional Bayesian inference to model the instrument noise and short duration gravitational
wave signals of arbitrary morphology.  The foundational principle behind this approach
is to allow the complexity of the model to automatically adapt according to the complexity of
 the data, following the motto ``model everything and let the data sort it out''.
This approach was implemented in the \BayesWave algorithm~\cite{Cornish:2014kda},
which models gravitational wave transients and noise transients as a collection of
continuous wavelets, and the  \BayesLine algorithm~\cite{Littenberg:2014oda}, which
models the power spectral density of the instrument noise using a smooth spline model
and a collection of Lorentzian lines. Since the  \BayesLine algorithm is a key component
and fully integrated in the \BayesWave algorithm, we will collectively refer to them as \BayesWave going forward.

The  \BayesWave algorithm has been used extensively in the analysis of data from the
LIGO~\cite{TheLIGOScientific:2014jea} and Virgo~\cite{TheVirgo:2014hva} gravitational wave detectors. Applications include template-free
reconstructions of gravitational wave signals, residual based tests of general relativity,
noise transient removal and power spectral density estimation for parameter estimation studies.
In the intervening years, the \BayesWave algorithm has undergone changes, with the
addition of new functionality and improvements in the sampling efficiency, and computational
cost. This paper serves as an update to \BayesWave, consistent with the publicly available
software at  \url{https://git.ligo.org/lscsoft/bayeswave}. Sec. \ref{sec:overview} serves as an overview of the
\BayesWave algorithm and use cases. Updates are organized by changes to the underlying data
model used by \BayesWave in Sec. \ref{sec:model}, and changes to the stochastic sampling
engine which improve convergence are described in Sec. \ref{sec:sampler}.
Sec. \ref{sec:post_processing} describes the inherent post-processing steps, while Sec.~\ref{sec:review}
discusses standard review tests for samplers. 
Sec. \ref{sec:conclusions} concludes by outlining future development and use-cases. 
Appendices \ref{sec:AppA}, \ref{sec:AppB}, \ref{sec:AppC} go into details about the
deployment, performance, and optimizations, respectively.

%%%%%%%%%%%%%%%%%%%%%%%%%%%%%%%%%%%%%%%%%%%%%%%%%%
\section{Overview of \BayesWave}
\label{sec:overview}
%%%%%%%%%%%%%%%%%%%%%%%%%%%%%%%%%%%%%%%%%%%%%%%%%%
The foundation of the \BayesWave algorithm is a model for the data from detector $I$: $d_I = h_I + g_I + n_I$, where $h_I$ is the detector's response to a gravitational wave signal, $g_I$ are non-Gaussian noise transients, or ``glitches" in the data, and $n_I$ is the random detector noise.  The transients $h$ and $g$ are modeled using coherent parameterized fits to the waveforms $h_I^M(\theta_h)$ and $g_I^M(\theta_{g,I})$, with superscript $M$ indicating that it is a model of the true signal, while $n_I$ is modeled statistically with a parameterized noise covariance matrix $C_I(\theta_n)$.  The model parameters are optimized using a Markov Chain Monte Carlo (MCMC) sampler of the posterior distribution function
\begin{equation}
p(\mathbf\theta \vert d) = \frac{p( d \vert \mathbf\theta)p(\mathbf\theta) }{p(d)}
\end{equation}
where the joint parameter set is $\mathbf\theta = [\theta_h,\theta_{g,I}, ...,\theta_{g,K},\theta_{n,I}, ...,\theta_{n,K}]$ for detectors $I, ..., K$, and the joint data are $\mathbf d = [d_I,...d_K]$ and similarly for the signal $\mathbf{h^M}$ and glitch $\mathbf{g^M}$ models.
The likelihood for the Fourier domain residual $\tilde{\mathbf r} = \tilde{\mathbf d} - \tilde{\mathbf{h}}^\mathbf{M} - \tilde{\mathbf{g}}^\mathbf{M}$, assuming that the remaining noise is Gaussian distributed, is
\begin{equation}\label{eq:pedantic_likelihood}
p(d|\mathbf\theta) = \frac{1}{ \det( \pi\mathbf{C})}e^{- \tilde{\mathbf r}^*  \mathbf C^{-1}  \tilde{\mathbf r} }
\end{equation}   
and $p(\mathbf\theta)$ and $p(d)$ are the prior and marginalized likelihood, respectively.

When data from multiple detectors are considered the noise is assumed to be independent between detectors so the full noise covariance matrix $\mathbf{C}$ is block diagonal. Furthermore, the variance of the noise is assumed to be constant over the observation period (stationary) of duration $T$, and thus each block of $\mathbf{C}$ is itself diagonal with the only non-zero elements being proportional to the variance $\langle n_{[i,j]} n_{[k,l]}^* \rangle = \frac{T}{2} S_{n,[i,j]} \delta_{i,k} \delta_{j,l}$, where $S_{n,[i,j]}$ is the noise power spectral density $S_n$ of detector $i$ in data sample $j$.  Under these assumptions, the likelihood in Eq.~\ref{eq:pedantic_likelihood} reduces to 
\begin{equation}\label{eq:like}
p(d|\mathbf\theta) = \prod_i\prod_j \frac{2}{\pi T  S_{n,[i,j]}}e^{- \frac{ 2|\tilde{r}_{[i,j]}|^2}{TS_{n,[i,j]}}} 
\end{equation}
again with indices $i$ spanning the data streams and $j$ the data samples. See Ref.~\cite{Romano:2016dpx} for details on the likelihood derivation for discretely sampled data in the Fourier domain.

As originally described in Ref. \cite{Cornish:2014kda}, both the signal and the glitch models are constructed from a linear combination of sine-Gaussian wavelets 
\begin{eqnarray}\label{mg}
\Psi(t; \vec{\lambda}) &=& A e^{(t-t_0)^2/\tau^2} \cos(2\pi f_0 (t-t_0) + \phi_0) \nonumber \\
\tilde\Psi(f; \vec{\lambda}) &=& \frac{\sqrt{\pi} A \tau}{2}  e^{ -\pi^2 \tau^2 (f-f_0)^2} \left( e^{ i(2\pi(f-f_0)t_0+\phi_0) } \right.\nonumber \\
&& \quad \left. + e^{-i(2\pi(f+f_0)t_0+\phi_0)e^{-Q^2f/f_0}} \right)
\end{eqnarray}
with $\vec{\lambda} \rightarrow (t_0,f_0,Q,A,\phi_0)$, where $t_0$ is the central time of the wavelet, $f_0$ is frequency at $t=t_0$, $Q$ is the wavelet quality factor (i.e. the number of cycles of the wavelet over one $e$-folding of the Gaussian envelope), $\tau=Q/2\pi f_0$, $A$ is the wavelet amplitude, and $\phi_0$ is the wavelet phase at $t=t_0$. The wavelets form a frame, not a basis, since they are not linearly independent. The glitch model wavelet parameters are independent in each detector, while the signal model wavelets are coherently projected on to each detector using a set of extrinsic parameters (see Sec.~\ref{polarization}).
In the standard configuration, \BayesWave assumes that the gravitational wave signal is elliptically polarized, as is the case for quasicircular, non-precessing compact binary coalescences (CBCs). In that case there are four extrinsic parameters: sky location angles specifying the right ascension $\alpha$ and declination $\delta$, the polarization angle $\psi$, and the ellipticity parameter $\epsilon$ which maps the $+$ polarization to the $\times$ polarization via $h_\times = \epsilon h_+ e^{i\pi/2}$. 
The total number of wavelets used in the glitch and signal model is marginalized over with a trans-dimensional MCMC, making the number of wavelets in each model a free parameter of the model.

\BayesWave was designed with a generic application programming interface for the wavelet functions. In principle, the sine-Gaussian wavelets are easily replaced by any other set of frame functions.  There is currently one alternative wavelet model available in \BayesWave. Reference~\cite{Millhouse:2018dgi} describes the inclusion of a \emph{chirplet} frame which are similar to the sine-Gaussian wavelets but modified with a constant frequency derivative $\dot{f}_0$. The added flexibility of the chirplet model to track rapidly changing frequency content of a signal allows for better fits to certain waveform morphologies.

The Gaussian noise model is handled by an independently developed MCMC algorithm, \BayesLine \cite{Littenberg:2014oda}, although the name \BayesWave has become synonymous for both algorithms.
In \BayesLine, the Gaussian noise model is decomposed into two components evident in LIGO and Virgo data--broad spectrum and gradually varying noise due to e.g., ground motion, thermal fluctuations in the optics, and shot noise from the interferometric sensing; and narrow band, high amplitude, spectral lines due to mechanical resonances in the detectors, the power supply at detector, and calibration lines intentionally added to the data. The broadband noise is modeled as a cubic splines interpolation between control points parameterized by their frequency and noise level, the number and location of which are adjustable by the sampler. The narrow band noise is modeled with a linear combination of Lorentzian-like functions parameterized by the central frequency, width, and line height.  Again, a trans-dimensional MCMC algorithm is used to marginalize over the number of Lorentzians used in the fit.
\BayesLine has been shown to outperform periodogram-based approaches for spectral estimation in LIGO-Virgo data because it only assumes the noise to be stationary over the interval of data being analyzed, as opposed to the interval of data needed for the periodogram~\cite{Chatziioannou:2019}.

The combined \BayesWave algorithm uses a blocked Gibbs sampler, alternating between updates to Gaussian noise model, the intrinsic  parameters which control the number and shape of the wavelet model, and (for the signal model) the extrinsic parameters which govern the coherent projection of the model onto the network of detectors.

To evaluate different hypothesis about the data, \BayesWave is run in a restricted setting only allowing certain models in the fit and then using thermodynamic integration~\cite{gelman1998} to compute the evidence for the model.  The standard workflow includes assessing a Gaussian noise-only model, a model containing Gaussian noise and the joint glitch model for all detectors, and the Gaussian noise plus signal model which requires \emph{at least one wavelet} to be coherently projected onto the network.  The Bayes factor between Gaussian noise plus signal model and the Gaussian noise plus glitch model is a robust detection statistic~\cite{Littenberg:2016}.

\subsection{Existing Proposals \& Priors}

\paragraph*{Priors:}
The \BayesWave default behavior is to use flat priors on all intrinsic parameters except the wavelet amplitude $A$, and for all extrinsic parameters.
For intrinsic parameters, the prior ranges cover: 
$t_0 \in U[t_{\rm min},t_{\rm max}]$ where by default $t_{\rm min}$ and $t_{\rm max}$ enclose a 1 s interval centered on the candidate GW event time, though the interval and location in the full data segment are adjustable by the user; 
$f_0 \in U[f_{\rm min},f_{\rm max}]$ where $f_{\rm min}$ is specified by the user and $f_{\rm max}$ is the Nyquist frequency determined from the user-requested sampling rate for the input data;
$Q \in U[0.1,40]$ by default but is adjustable by the user; 
$\phi_0 \in U[0,2\pi]$.
For signal-model extrinsic parameters, the ranges
$\alpha \in U[0,2\pi]$;
$\sin\delta \in U[-1,1]$ which, combined with the prior on $\alpha$ make the joint prior uniform on the sky;
$\psi \in U[0,\pi]$; 
$\epsilon \in U[-1,1]$;
and an overall phase applied to all wavelets in the signal model $\varphi \in U[0,2\pi]$. This overall phase shift is degenerate
with a simultaneous shift of all wavelet phases $\phi_0$ by the same amount, but the explicit inclusion of $\varphi$ aids the sampler convergence.

The amplitude prior is based on the signal-to-noise ratio, $\SNR$, of each wavelet and is designed to suppress low amplitude wavelets which will not contribute to the likelihood while also not biasing the amplitude recovery of high $\SNR$ signals. 
The priors take slightly different functional form for the glitch and signal model to account for the fact that  high amplitude glitches ($\SNR>100$) are not uncommon in the data whereas GW signals do not reach such levels at the current detector sensitivities.  More detail on the amplitude prior is provided in Sec~\ref{sec:amp_prior}.

By default \BayesWave also uses flat priors on the number of wavelets $D \in [D_{\rm min},100]$ though the maximum number of wavelets is adjustable by the user, with $D_{\rm min}=1(0)$ for the signal (glitch) model.  Note that nested within the glitch model is the Gaussian noise model, however it is sometimes worth evaluating the Gaussian-noise only model by itself as the posterior weight of the glitch model in the zero wavelet case may be impractically small given the number of posterior samples in the chain.

\paragraph*{Proposals:}
\BayesWave uses a mixture of different proposal distributions for generating trial parameters. These include fair draws from the prior to ensure efficient sampling of the full parameter space for the high temperature chains. Draws from the prior are used for within-model and transdimensional proposed moves. \BayesWave also relies on custom-made proposals that leverage what is known, or can easily be inferred, from the data or model. In addition new developments described in Sec. \ref{sec:sampler}, the \BayesWave sampler particularly benefits from proposals along eigenvectors, scaled by the eigenvalues, of the Fisher Information Matrix approximation to the inverse covariance matrix $C_{ij}^{-1}\sim\Gamma_{ij}\equiv (\frac{\partial \Psi }{\partial \lambda_i} | \frac{\partial \Psi}{\partial \lambda_j} )$.  The elements of the Fisher matrix for the wavelets are known analytically, while matrix elements for the extrinsic parameters of the signal model are computed numerically.  

\BayesWave also uses proposals that encourage placing new wavelets near in time-frequency to existing groups of wavelets.  The proposal is built by summing Gaussians at existing wavelet locations in such a way that regions in the time-frequency plane near to, but not overlapping, with existing wavelets are preferentially tried by the sampler. All of the aforementioned proposals are explained in Ref.~\cite{Cornish:2014kda}.

One final proposal worth noting here takes advantage of a near degeneracy between the reference time and reference phase of a wavelet, particularly for wavelets with a high $Q$ parameter.  The degeneracy arrises because the wavelets can be time-shifted by a cycle and still match a high-$Q$ feature in the data by compensating by adjusting the wavelet phase.  This proposal is described in detail in Ref.~\cite{becsy:2017}.

\subsection{Use cases:}
The flexibility of the \BayesWave signal, glitch, and noise models to adapt to features in the data, 
coupled with the restraint on the models applied by comparing evidences, 
have made the algorithm well-suited to a broad range of LIGO-Virgo analyses as a tool to 
study both the signals and the noise in the data. On the search side, \BayesWave is part of
a hierarchical detection pipeline for generic short-duration gravitational wave transients 
(i.e. ``bursts'')~\cite{Kanner:2016,TheLIGOScientific:2016uux,Abbott:2019prv}. Regarding glitches
in data that contain a candidate signal,
\BayesWave was adapted as a data cleaning tool to subtract the noise transient 
that overlapped with GW170817~\cite{TheLIGOScientific:2017qsa}, while preserving the fidelity
of the underlying signal analysis~\cite{Pankow:2018qpo}. Extensive glitch-subtraction was subsequently
part of the GWTC-2 catalog~\cite{Abbott:2020niy}.
On the noise side, \BayesWave is used as a spectral estimation tool that provides the noise model 
to the template-based parameter estimation follow up of compact binary signals~\cite{Chatziioannou:2019,LIGOScientific:2018mvr}.
 
Regarding analysis of gravitational wave signals, \BayesWave has been used to study unmodeled or poorly-modeled sources, such as
the post-merger emission from neutron star binaries~\cite{Chatziioannou:2017ixj,Abbott:2018wiz,Torres-Rivas:2018svp}, 
eccentric black hole binaries \cite{dalya:2020}, post-merger echoes from black hole binaries~\cite{Tsang:2018uie,Tsang:2019zra},
and supernova~\cite{Gill:2018hxg}. 
Additionally, it has contributed waveform reconstructions and inferences of candidate GW transients without the use of templates. 
\BayesWave's generic reconstructions are compared to the template-based reconstructions as part of event validation 
studies~\cite{Abbott:2016blz,TheLIGOScientific:2016wfe,LIGOScientific:2018mvr,Ghonge:2020suv}.
 Related to this, \BayesWave has been used to search for excess power in residual data obtained after the subtraction of a template-based 
 point estimate to signals, thus testing how well the physically motivated waveform models match the data; this analysis 
 has been interpreted as a 
 model-agnostic test of General Relativity~\cite{TheLIGOScientific:2016src,LIGOScientific:2019fpa,Abbott:2020jks}.

%%%%%%%%%%%%%%%%%%%%%%%%%%%%%%%%%%%%%%%%%%%%%%%%%%
\section{Model Extensions}
\label{sec:model}
%%%%%%%%%%%%%%%%%%%%%%%%%%%%%%%%%%%%%%%%%%%%%%%%%%

In this section we describe the new \BayesWave model capabilities, including changes in the signal and glitch models, as well as new 
supported priors.

%%%%%%%%%%%%%%%%%%%%%%%%%%%%%%%%%%%%%%%%%%%%%%%%%%
\subsection{Signal polarization}
\label{polarization}
%%%%%%%%%%%%%%%%%%%%%%%%%%%%%%%%%%%%%%%%%%%%%%%%%%

In General Relativity, GWs contain two polarization modes, colloquially referred to as plus and cross. GW detectors respond differently to each of these modes, encoded in the
detector antenna pattern functions $F_{\times}(\Omega,\psi),  F_{+}(\Omega,\psi)$, where $\vec\Omega\rightarrow(\alpha,\delta)$ describes the sky location of the source and $\psi$ is the polarization angle. The response of a detector to an impinging signal can then be expressed as
\begin{equation}
h_{I}(f) = (F_{\times}(\Omega,\psi)h_{\times}(f)+F_{+}(\Omega,\psi)h_{+}(f))e^{2\pi i f \Delta t(\Omega)}\label{hresponse},
\end{equation}
where $h_{I}(f)$ is the interferometric response, $\Delta t(\Omega)$ is the light travel time from a fiducial reference location to the detector, 
and $h_{\times}(f),h_{+}(f)$ are the cross and plus signal respectively expressed at the reference location. \BayesWave uses an arbitrarily chosen detector as the reference location.

The projection from the reference location to each individual detector in the network, Eq.~\eqref{hresponse}, contains not only the two GW polarization modes, but also the sky location and orientation. This means
that with two or three detectors available, the problem of extracting GW polarizations from the observed detector response might be under-constrained. For this reason, the original signal model in  \BayesWave
restricted the polarization content of the signal to the case of elliptical polarization
\begin{eqnarray}
\hplus &=& \sum_n\wavelet(f;t_0^n,f_0^n,Q^n,\mathcal{A}^n,\phi_0^n), \nonumber \\
\hcross &=& i\epsilon h_+,
\end{eqnarray}
where $\epsilon$ is the \emph{ellipticity} parameter encoding the degree of elliptical polarization. For $\epsilon=0$, $\hcross=0$ and the signal is linearly polarized; if $\epsilon=1$ the signal is circularly polarized.
This assumption of elliptical polarization does not hold for complicated CBC signals whose polarization content changes with time, such as spin-precessing signals or signals with a strong higher-order modes content.  Additionally, generic bursts of GWs are not expected to possess any special polarization content. 
Previous studies have shown that restricting the signal model to elliptical polarization is sub-optimal for detecting unpolarized signals such as white noise bursts~\cite{becsy:2017}.

In order to relax the elliptical polarization constraint, we generalize the signal model in  \BayesWave to
\begin{eqnarray}
\hplus &=& \sum_n\wavelet(f;t_0^n,f_0^n,Q^n,\mathcal{A}^{n,+},\phi_0^{n,+}), \nonumber \\
\hcross &=& \sum_n\wavelet(f;t_0^n,f_0^n,Q^n,\mathcal{A}^{n,\times},\phi_0^{n,\times}),
\end{eqnarray}
while setting $\psi=0$ in Eq.~\eqref{hresponse}. This generic polarization model assumes that each polarization state can be expressed as a sum of the same number $n$ of wavelets that have the same quality
factor $Q^n$, central time $t^n_0$, and central frequency $f^n_0$, but differ in amplitude and phase. We argue that we can restrict the plus and cross wavelets to the same $(Q,t_0,f_0)$ without loss of generality. From
Eq.~\eqref{hresponse} it is clear that the time-frequency content of the GW signal is independent of the detector network, and only its amplitude and phase are modified by the process of projecting it from  the reference location onto a detector network. This has the added benefit of avoiding pathological solutions where all of $\hplus$ is in one detector and all of $\hcross$, making it degenerate with the glitch model.

We showcase the generic polarization model by analyzing simulated CBC signals observed by a network of Hanford, Livingston and Virgo (HLV) GW detectors at 
design sensitivity with a network signal to noise ratio $\SNR=100$. 
To simulate the observed data we use the waveform model IMRPhenomPv2~\cite{Hannam:2013oca} and assume a zero noise realization. The component masses are
set to $m_1=20M_{\odot},m_2=5M_{\odot}$, as unequal masses are known to maximize the effect of precession, 
and hence deviation from elliptical polarization~\cite{Fairhurst:2019vut,Pratten:2020igi}.  
Besides the system's mass ratio, the binary inclination and the amount of in-plane spin also affect the degree to which a signal is precession-modulated. We employ 
two values of the inclination angle between the line of sight and the orbital angular momentum $\iota=\{45^\circ,90^\circ\}$, defined at $f=16$Hz. The inclination angle 
evolves under spin-precession, so an originally edge-on system ($\iota=90^\circ$) will not remain in this configuration. The in-plane spin is commonly characterized through
the $\chi_p$ parameter~\cite{Schmidt:2014iyl} and we inject signals with $\chi_p=\{0, 0.52, 0.98\}$, again defined at $16$Hz. 
The first case corresponds to a spin-aligned system, studied for reference. The value of
$\chi_p$ also evolves under spin-precession, and it is not directly related to how prominent precessional modulations are in the observed signal~\cite{Fairhurst:2019vut}.

\begin{figure}[]
    \parbox{\hsize}{
    \includegraphics[width=\hsize]{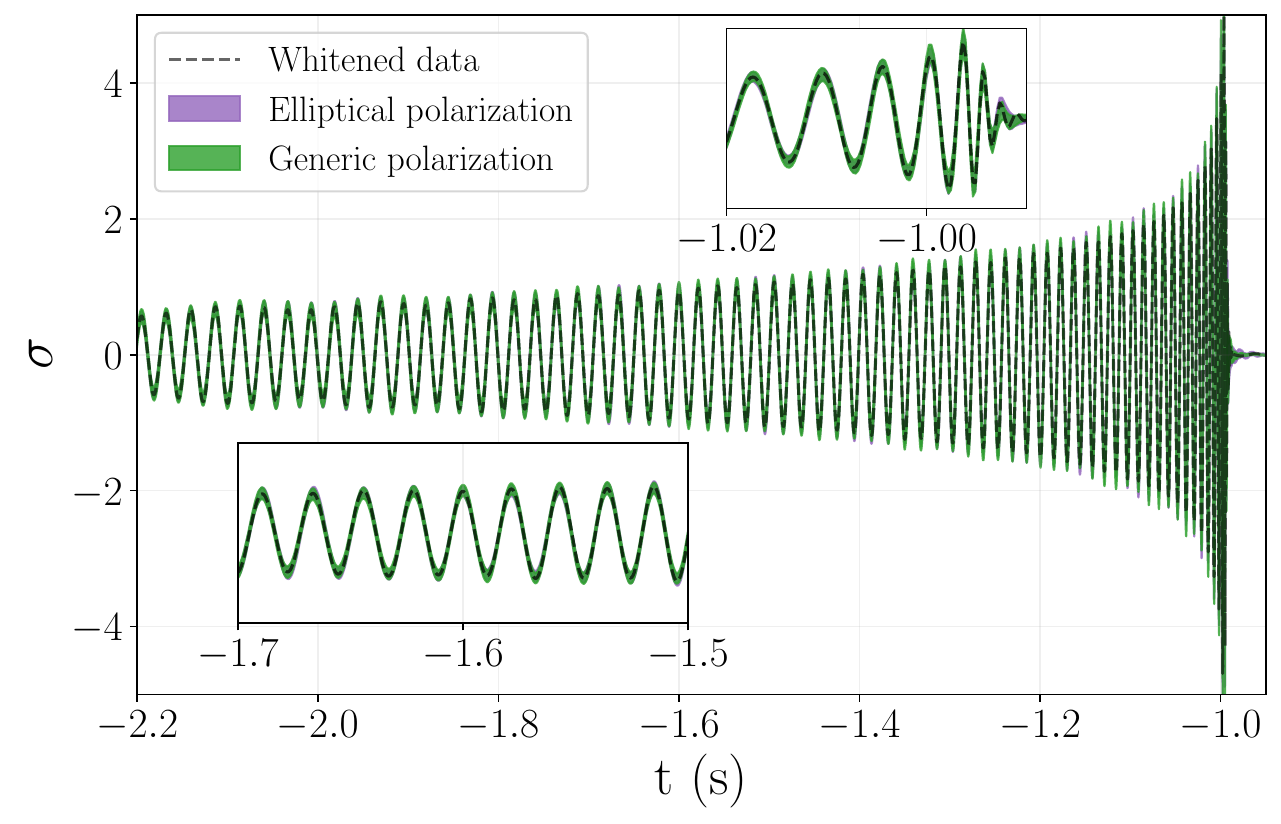}\\[-2ex]
    }\\
    \parbox{\hsize}{
    \includegraphics[width=\hsize]{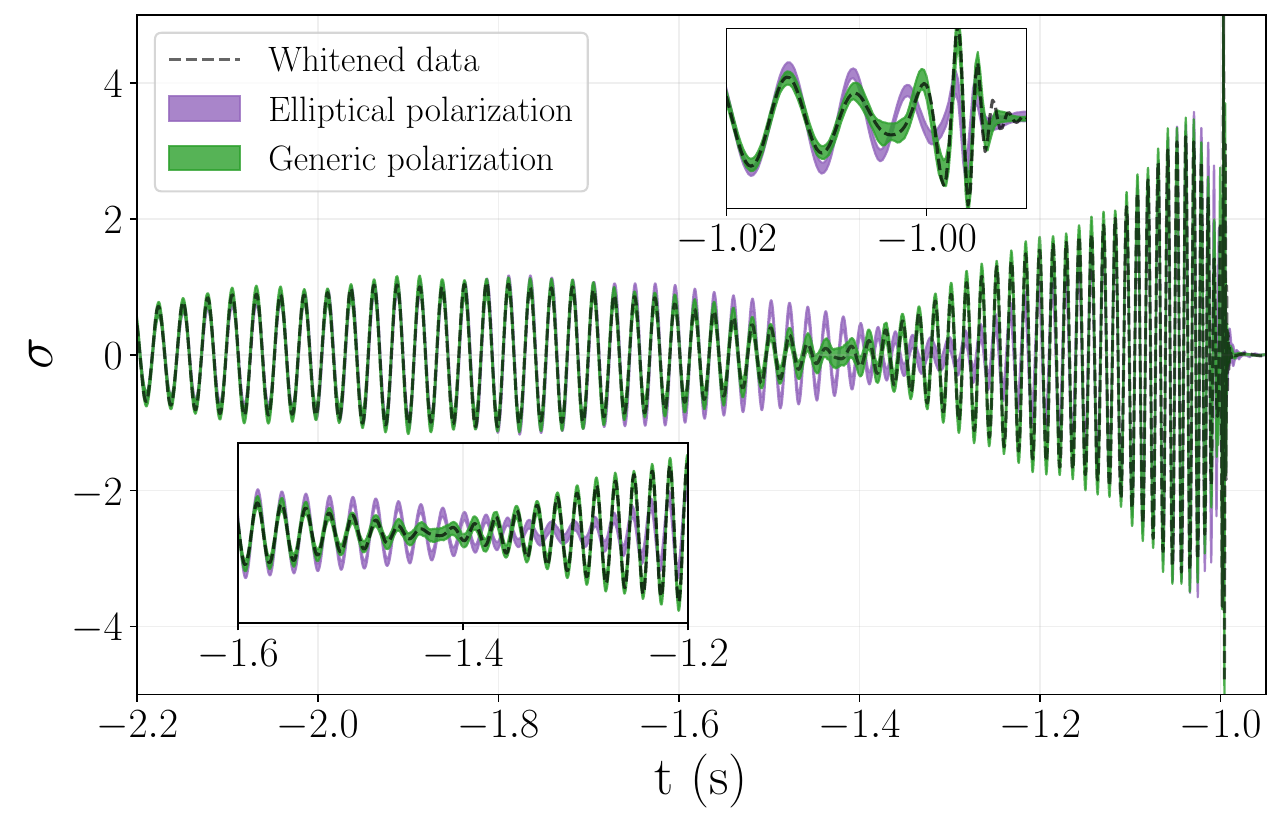}\\[-2ex]
    }\\
    \parbox{\hsize}{
    \includegraphics[width=\hsize]{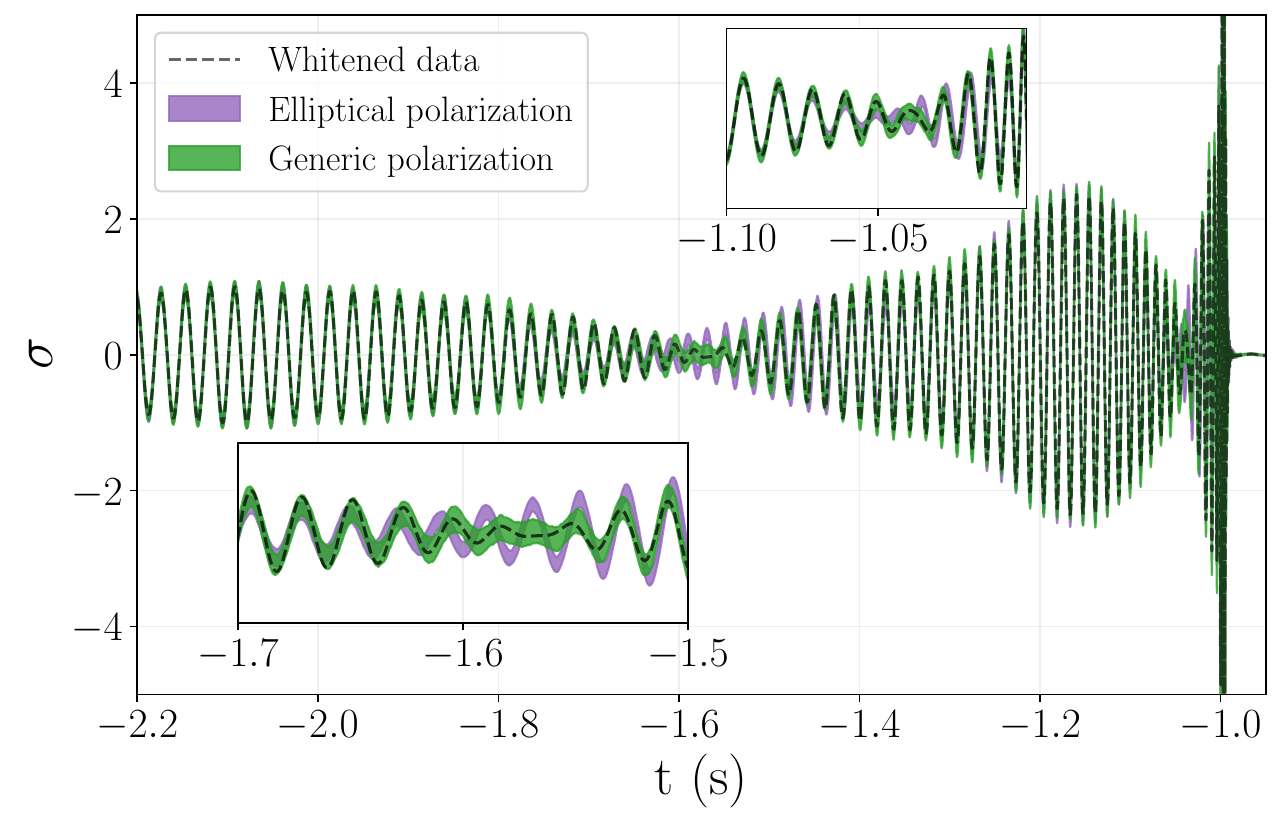}\\[-2ex]
    }
    \caption{Whitened time-domain reconstructions of injected GW signals. We plot the injected data in grey dashed lines, where no additional noise realization has been added. 
    Purple (green) shaded regions show the 90\% credible interval for the reconstruction when assuming an elliptical (generic) polarization content for the signal. From top
    to bottom we have signals with network signal to noise ratio $\SNR=100$ and with $(\chi_p=0,\iota=45^{\circ})$, $(\chi_p=0.52,\iota=45^{\circ})$, $(\chi_p=0.98,\iota=90^{\circ})$.}
    \label{fig:PolRec}
\end{figure}

We show reconstructions of the observed data in the Livingston detector in Fig.~\ref{fig:PolRec} for different values of $\chi_p$. Shaded regions show the
90\% credible interval for the reconstruction when assuming an elliptical polarization (purple) and a generic polarization (green). The top panel contains a signal with 
$\chi_p=0$, which exhibits no precessional modulations. Both analyses reconstruct the signal similarly well. The middle and bottom panels show the results for
$\chi_p=0.52$ and $\chi_p=0.98$ respectively. In both cases we find that the  \BayesWave analysis that allows for a generic signal polarization does a better job of reconstructing
the injected signal. Despite this improvement, the elliptical polarization analysis is still able to reproduce the spin-induced amplitude modulation to some extent, suggesting 
that signals of extreme $\SNR$ and precession are needed before the elliptical polarization approximation results in considerably deteriorated inference. Interestingly,
we also find that the elliptical polarization analysis performs better for the $\chi_p=0.98$ signal than the $\chi_p=0.52$ one, again suggesting that $\chi_p$
might not be a suitable parameter to quantify the amount of spin-precession present~\cite{Fairhurst:2019vut}. 

\begin{figure*}[]
    \parbox{0.5\hsize}{
    \includegraphics[width=\hsize]{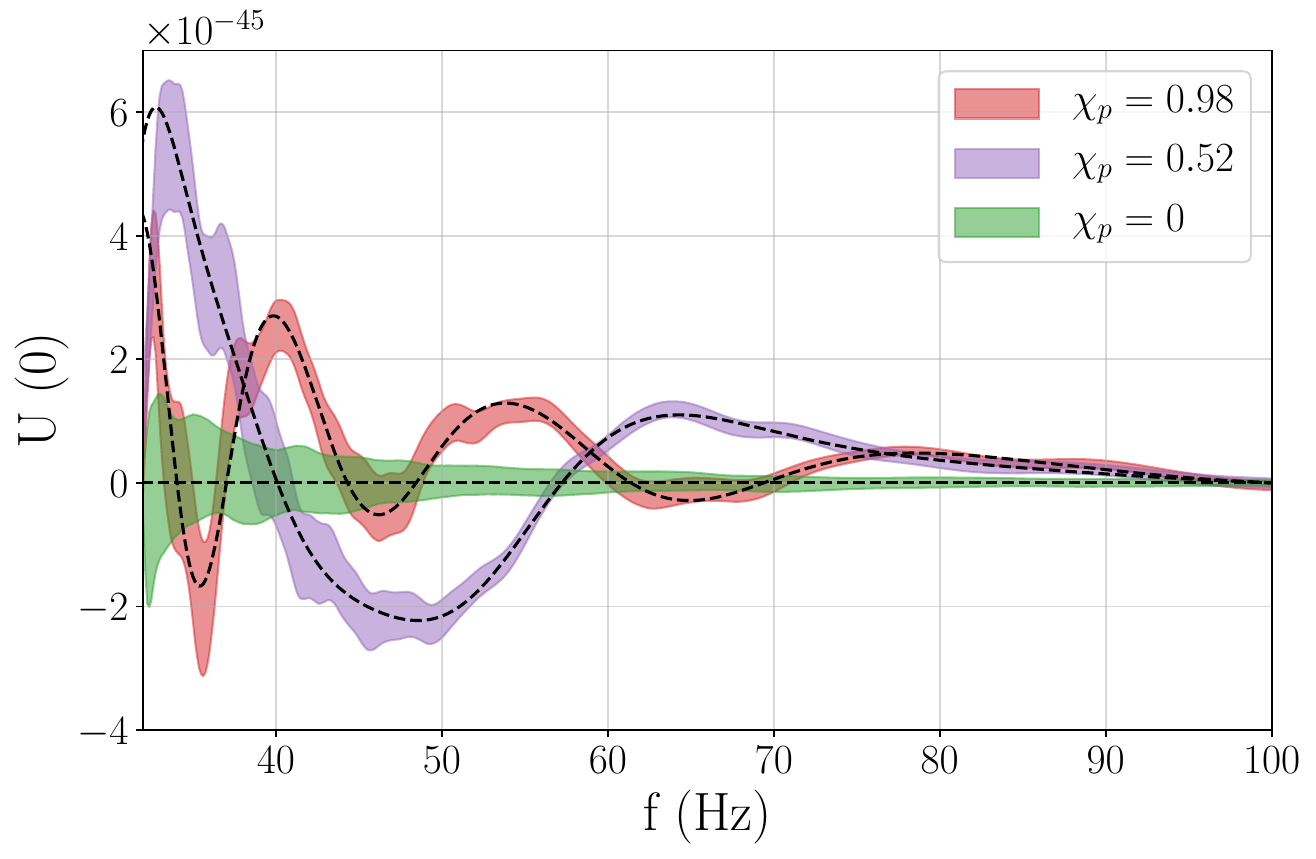}\\
    }\parbox{0.5\hsize}{
    \includegraphics[width=\hsize]{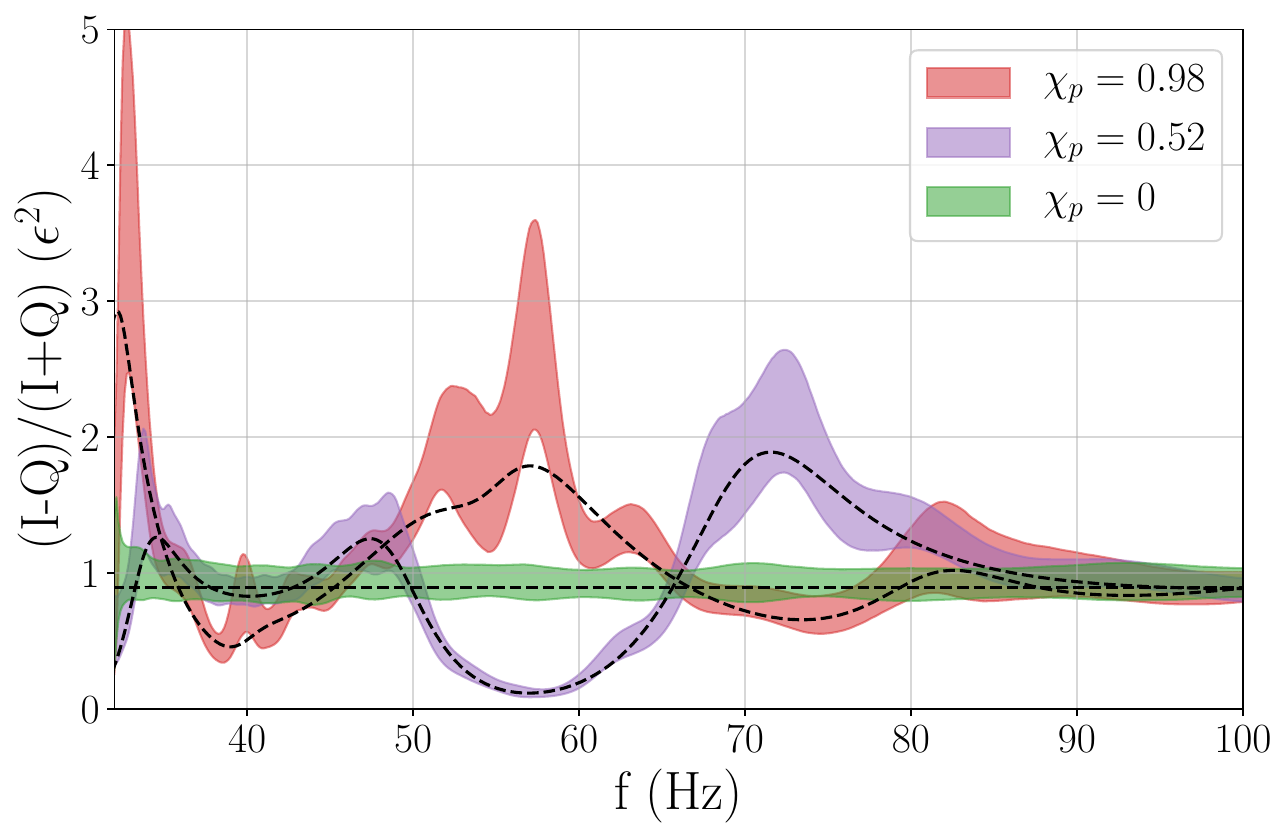}\\
    }\\
    \parbox{0.5\hsize}{
    \includegraphics[width=\hsize]{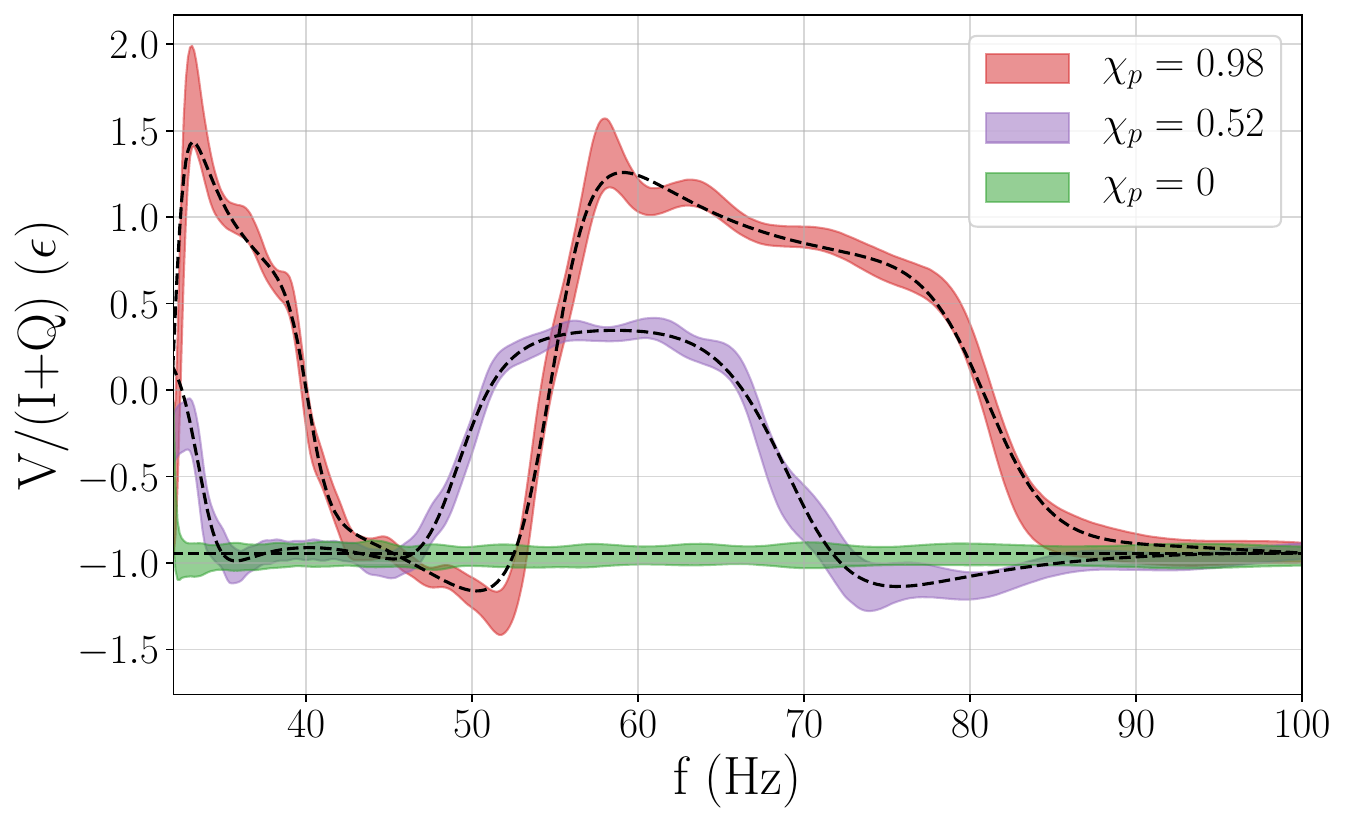}\\
    }\parbox{0.5\hsize}{
    \includegraphics[width=\hsize]{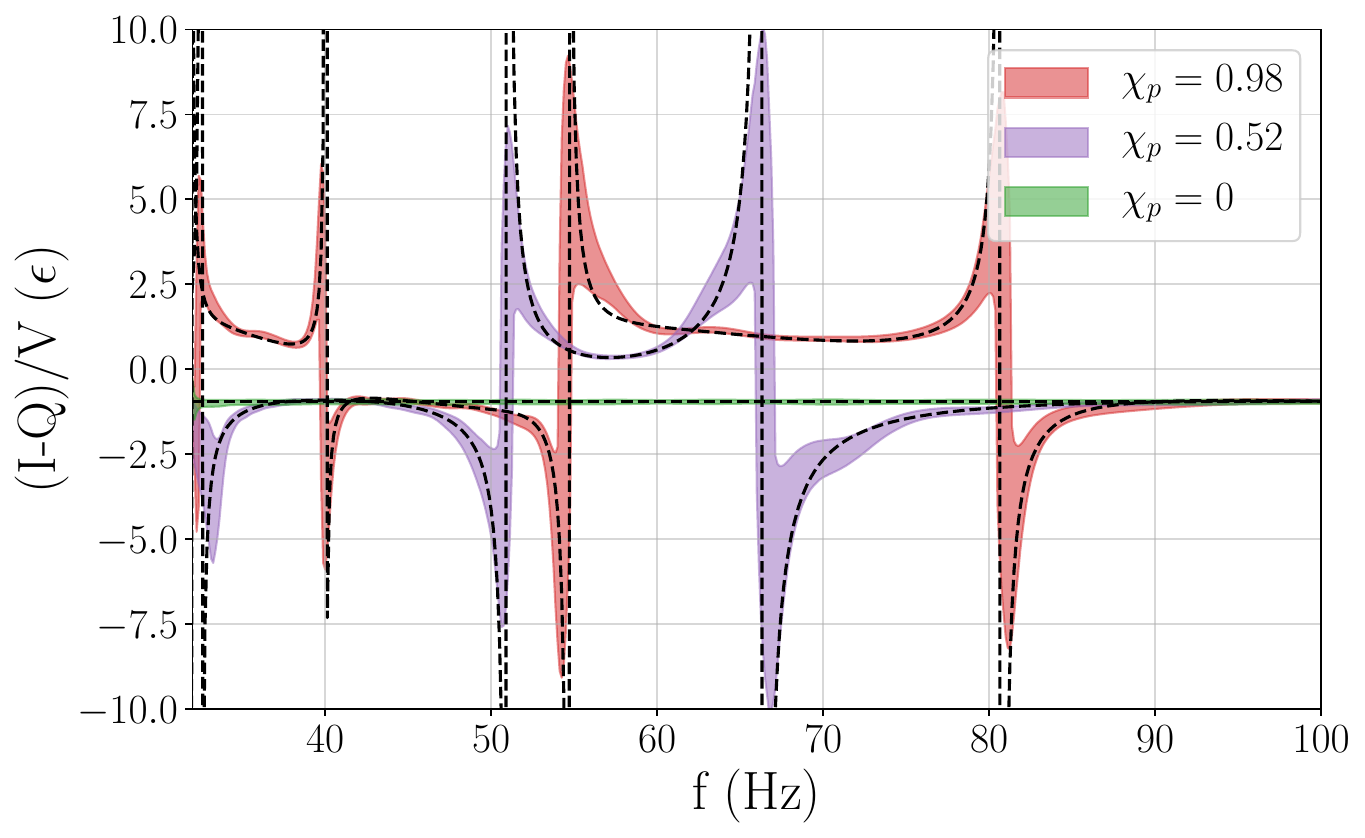}\\
    }
    \caption{Stokes parameter combinations as a function of frequency for the three injections at $\SNR=100$ and $\iota=45^{\circ}$. Dashed lines show the injected values and
    shaded regions show the 90\% credible interval obtained under the generic polarization analysis. The parenthesis on the y axis label indicates the constant value each 
    Stokes combination assumes under elliptical polarization.
          }
    \label{fig:Stokes}
\end{figure*}

Besides more faithful signal reconstruction, the generic polarization analysis also allows us to infer the polarization content of the observe signal. 
We employ the usual Stokes parameters~\cite{Romano:2016dpx}, defined as
\begin{align}
U & = \tilde{h}_+ \tilde{h}_{\times}^* + \tilde{h}_{\times} \tilde{h}_+^*,\\
V & = i(\tilde{h}_+ \tilde{h}_{\times}^* - \tilde{h}_{\times} \tilde{h}_+^*),\\
I & = |\tilde{h}_+|^2+|\tilde{h}_{\times}|^2,\\
Q & = |\tilde{h}_+|^2-|\tilde{h}_{\times}|^2,
\end{align}
which here are to be understood as being a function of the GW frequency.
For an elliptically polarized signal with ellipticity $\epsilon$ these reduce to $U\sim 0, (I-Q)/(I+Q)\sim \epsilon^2, V/(I+Q)\sim \epsilon, (I-Q)/V\sim \epsilon$. 
Figure~\ref{fig:Stokes} shows these combinations as a function of frequency for the three values of $\chi_p$ studied and $\iota=45^{\circ}$. The dashed black lines show the injected values computed directly from the simulated signal, while shaded regions show the 90\% credible interval for the reconstruction when employing the  generic polarization analysis. In all cases the generic polarization model is able to reconstruct the stokes parameters and their frequency evolution, suggesting that the generic analysis can be used to reconstruct the polarization content of a detected signal in a morphology-agnostic way.

%%%%%%%%%%%%%%%%%%%%%%%%%%%%%%%%%%%%%%%%%%%%%%%%%%
\subsection{Signal plus Glitch Model}
\label{signalandglitch}
%%%%%%%%%%%%%%%%%%%%%%%%%%%%%%%%%%%%%%%%%%%%%%%%%%

In the original version of  \BayesWave the signal and glitch models were disjoint hypotheses to be tested, where the former assumed that all wavelets in the model were coherent across the detector network, and the latter assumed that all wavelets were independent in each detector. 
We have now added a joint hypothesis where the data can contain both a coherent signal \emph{and} additional glitches in any detector's data (S+G).
The new model is made possible by the improved mixing of the Markov chain from the changes described in Section~\ref{sec:sampler}.

The joint model is of particular value when trying to identify and mitigate glitches that occur near a candidate GW signal, in which case the glitch model can be used to remove the excess noise from the data. Using \BayesWave to remove glitches from data near GW candidates was first used in the analysis of GW170817~\cite{TheLIGOScientific:2017qsa,Pankow:2018qpo}, when the part of the GW signal detectable by \BayesWave was sufficiently far from the glitch time that there was no concern of the wavelet model also removing some signal power.  
Using the S+G model improves the ``safety'' of the glitch subtraction as any coherent features in the data will be picked up by the signal model and only the excess power independent in either detector will be fit and removed by the glitch model.

\begin{figure*}[htbp]
\includegraphics[width=\textwidth]{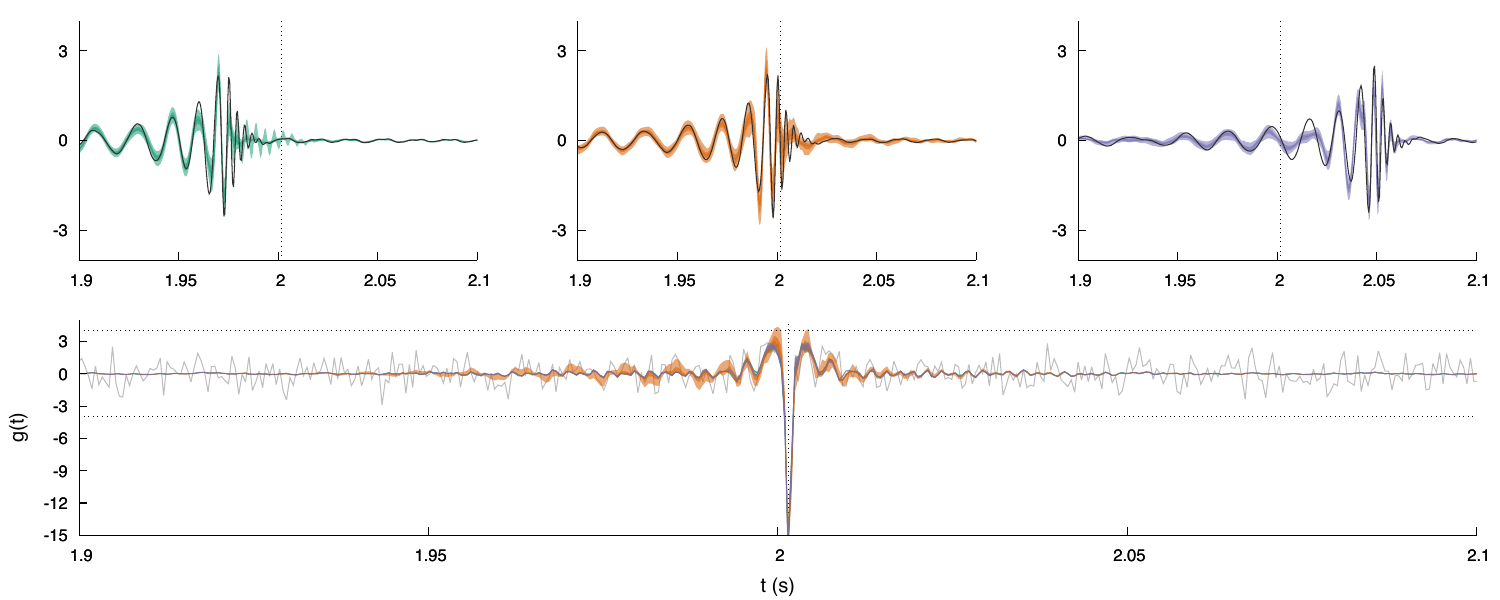}
\caption{Demonstration of the joint S+G analysis of data containing a noise transient in one detector and a coherent BBH signal in both data streams.  The top panels show credible intervals for the whitened waveform reconstructions (colored) the true waveform (black) for three different injections with merger time just before, coincident with, and after the glitch, indicated by the vertical dashed line.  The bottom panel shows the whitened data (gray) and the glitch reconstructions from each injection using the same color scheme as the top panel. The horizontal dashed lines indicate the vertical scale plotted in the top panel.  The joint model accurately separates the GW signal from the glitch despite the degree to which the two features in the data overlap in time and frequency. The test was performed on O1 data containing a common glitch type in one detector and otherwise clean in the other. The simulated GW events had parameters similar to GW150914 and were added to the data before analysis with \BayesWave.}
\label{fig:signal_plus_glitch}
\end{figure*}

Figure~\ref{fig:signal_plus_glitch} demonstrates the S+G model on data from the first observing run (O1) containing a common glitch type in one detector.  
In this example the glitch was in the Hanford detector, but the glitch type is common in both Hanford and Livingston~\cite{Cabero_2019}.
A simulated BBH signal with parameters similar to those of GW150914 was added to the data with merger time just before, coincident, and just after the glitch time. \BayesWave was used to process the data with the BBH signal added using the S+G model, and we compare the signal reconstructions (top panel) and the glitch reconstructions (bottom panel) for each injection. In each case the signal and glitch recovery is self consistent, regardless of how much the glitch and signal overlap in time. The glitch reconstructions are further compared to a glitch-only analysis of the original data with no signal added (dark gray) again showing that the glitch recovery, and therefore the subsequent glitch subtraction, is robust. 

\begin{figure}
\includegraphics[width=0.5\textwidth]{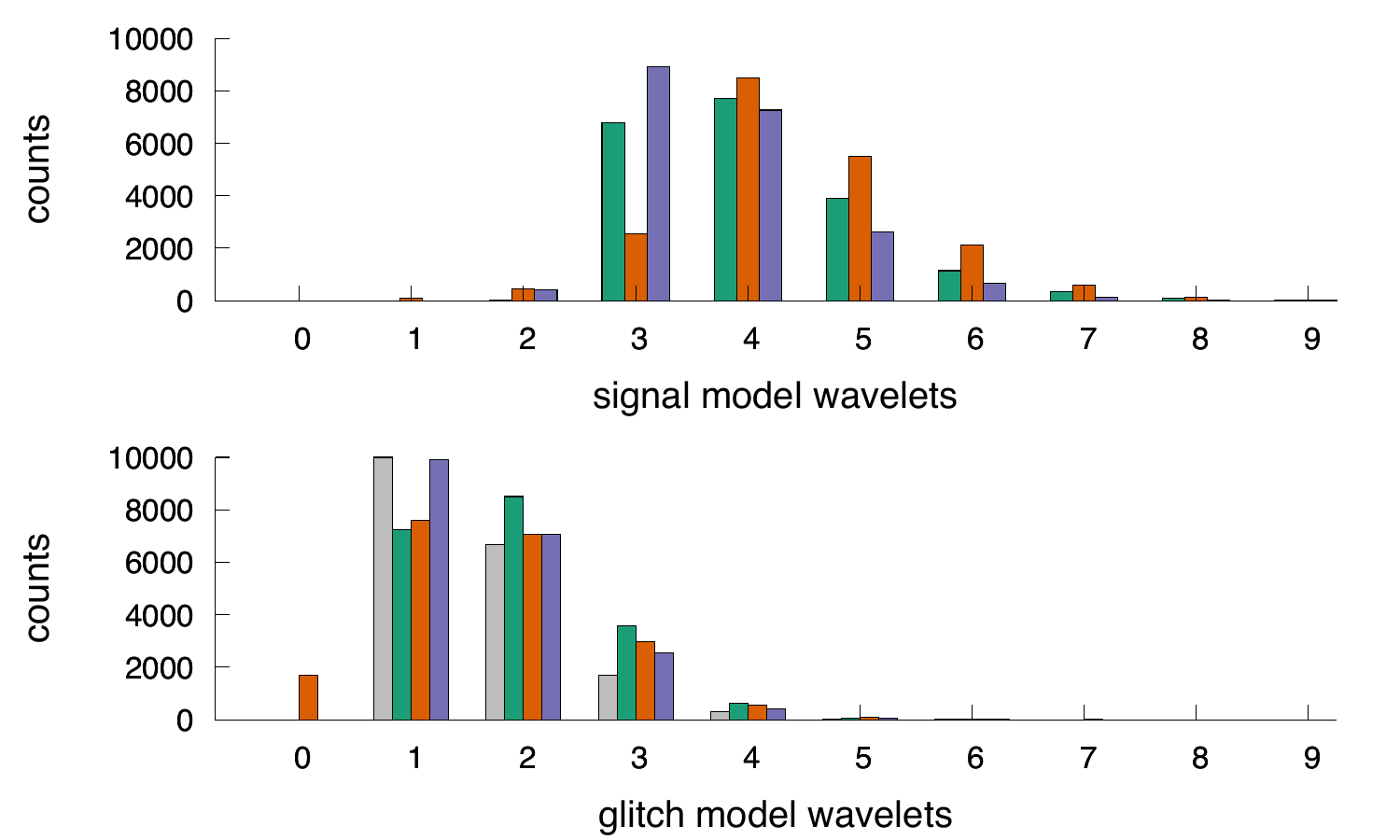}
\caption{Histograms of the total number of signal model wavelets and glitch model wavelets for the signal+glitch model analyses shown in Figure.~\ref{fig:signal_plus_glitch}. The color coding (green, orange, purple) correspond to the three signal injections show in the upper panel of Figure.~\ref{fig:signal_plus_glitch}. The additional grey colored bars are for when no signal is injected.}
\label{fig:signal_plus_glitch_dimension}
\end{figure}

Histograms of the number of wavelets used by the signal and glitch model for each each analysis are shown in Figure~\ref{fig:signal_plus_glitch_dimension}. The distribution of signal and glitch wavelets used is largely the same, no matter where the signal is injected relative to the glitch. The case where no signal is injected, indicated by gray bars in Figure~\ref{fig:signal_plus_glitch_dimension}, shows the same distribution of glitch model wavelets, but essentially no support for adding signal wavelets.

%%%%%%%%%%%%%%%%%%%%%%%%%%%%%%%%%%%%%%%%%%%%%%%%%%
%\subsection{Amplitude and Wavelet Number Prior}
\subsection{Updated Priors}
\label{sec:newpriors}

\subsubsection{Wavelet amplitude prior}\label{sec:amp_prior}
In both the signal and the glitch model, the prior on the amplitudes of the individual wavelets is actually given by a prior on the signal to noise ratio of an individual wavelet.
For Morlet-Gabor wavelets the signal to noise ratio is estimated as
\begin{equation}
  \SNR^2 \simeq \frac{A^2 Q}{2\sqrt{2\pi}f_0 S_n(f_0)},
\end{equation}
where $A$, $Q$, $f_0$ are the time, quality factor, and central frequency of the wavelet respectively, and $S_n(f_0)$ is the one-sided noise power spectral density at $f_0$.

\begin{figure}[h]
\begin{center}
\includegraphics[width=0.5\textwidth]{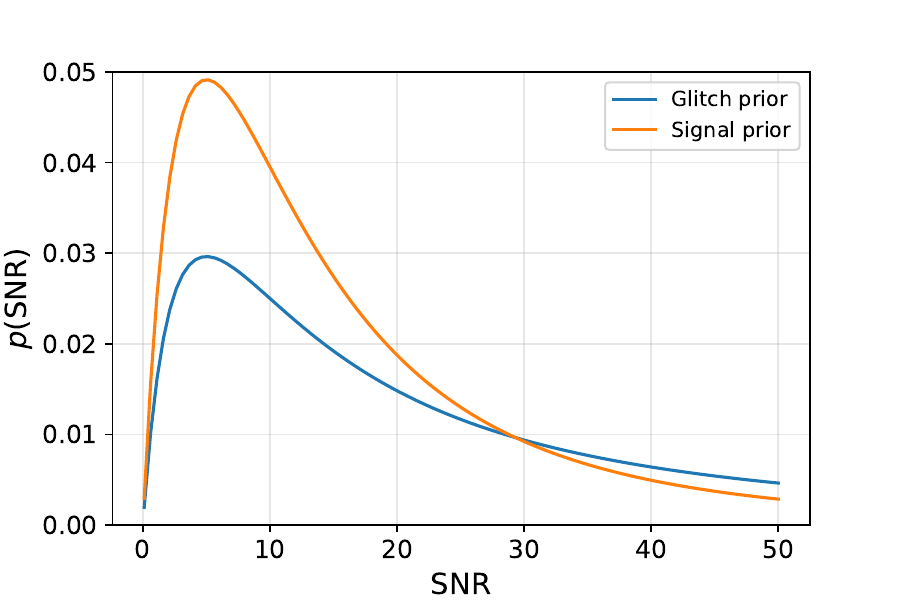}
\caption{Prior probability distribution on the signal-to-noise of individual wavelets, for both the signal and glitch model.  In this example both priors peak at $\SNR_*=5$.} %\mm{Does this plot need to be remade with SNR replaced with $\rho$?}}
\label{Fig:SNRprior}
\end{center}
\end{figure}

Because for both astrophysical signals and instrumental glitches we expect to get many more low $\SNR$ events than loud events, we formulate priors that peak at a given $\SNR_*$, and drop off at large and small $\SNR$. Having the prior go to zero at low $\SNR$ helps with convergence:  low amplitude wavelets have little effect on the likelihood and are disfavored by the natural parsimony of Bayesian inference. They eventually get discarded from the model, but it can take many iterations to shake them off. By shaping the prior to additionally disfavor low amplitude wavelets the convergence is accelerated. These convergence considerations apply equally to the signal and glitch models, and consequently, the same $\SNR_*$ value is used for both models.  Values between $\SNR_* = 3$ and $\SNR_* = 10$ have been found to yield similar results. The default value is set at $\SNR_* = 5$.
For the glitch model the prior is
\begin{equation}
  p(\SNR|\mathrm{glitch}) = \frac{\SNR}{2\SNR_*^2\left(1+\frac{\SNR}{2\SNR_*}\right)^3}.
  \label{eq:GlitchAmpPrior}
\end{equation}
For the signal model, the prior is
\begin{equation}
   p(\SNR|\mathrm{signal}) = \frac{3\SNR}{4\SNR_*^2\left(1+\frac{\SNR}{4\SNR_*}\right)^5}.
\end{equation}
The signal prior drops off as $\SNR^{-4}$, as expected from the distribution astrophysical sources. The glitch prior has been designed to have a heavier tail at large $\SNR$, since we expect to have more loud instrumental glitches than loud astrophysical signals.  The priors on $\SNR$ for the signal and glitch models are shown in Fig.~\ref{Fig:SNRprior}

\subsubsection{Wavelet dimension prior}
In the original  \BayesWave configuration, the prior on the number of wavelets ($N_w$) was flat.  After LIGO's first observing run, the distribution of wavelets in real data was used to develop a new prior.  Using an analytic fit to the 500 most significant background triggers for the \BayesWave's unmodeled transient search of O1 data~\cite{PhysRevD.95.042003}, the wavelet distribution was empirically modeled as
\begin{equation}
  p(N_w) = \frac{4\sqrt{3}N_w}{2\pi b^2(3+\frac{N_w}{b})^4}
\end{equation}
where $b=2.9$.

\subsubsection{Sky position}
For the extrinsic parameters, a feature was introduced to fix the sky location of the GW source to a known value.  This is used for cases where there is a known electromagnetic counter part to the GW signal, such as the kilonova associated with GW170817.  Using a fixed sky location speeds up run times, and can improve waveform reconstructions. 
%\mm{The fixed sky location can also be used to proved BFs between different polarisations, should I mention that here or is it out of scope?}
%%%%%%%%%%%%%%%%%%%%%%%%%%%%%%%%%%%%%%%%%%%%%%%%%%

%%%%%%%%%%%%%%%%%%%%%%%%%%%%%%%%%%%%%%%%%%%%%%%%%%
\section{Sampler updates}
\label{sec:sampler}
%%%%%%%%%%%%%%%%%%%%%%%%%%%%%%%%%%%%%%%%%%%%%%%%%%

In addition to changes made to the underlying data models tested by \BayesWave, there has been a similar scale of development to improve the efficiency of the pipeline.  
Improved performance can be found in two obvious places for an MCMC algorithm like \BayesWave -- shortening the ``burn in'' time when the sampler is locating the most likely modes in the distribution, and increasing the sampling efficiency, thereby decreasing the autocorrelation length, of the chain samples.

Satisfying both goals at once, we have continued developing customized proposal distributions used by the MCMC sampler.  
In general, proposal distributions which are good approximations to the target distribution will locate the important modes of the posterior more rapidly, sample the distribution with a smaller autocorrelation length.  
Below we describe proposals designed to leverage domain knowledge acquired either from a theoretical basis of how the likelihood function depends on the parameters, or proposals built from the input data itself.

In service of improving convergence time, we also describe a new model initialization procedure to get the sampler in a good starting position for the many components of the model, particularly in the case where large amplitude glitches bias the initial estimate of the noise spectrum which would then require a large number of samples to reduce down to the most parsimonious fit to the data.

\subsection{Sky Location Proposal}

The sky location and source orientation are poorly constrained for a two-detector network, often exhibiting multiple posterior
modes that are a challenge to sample. These degeneracies are reduced with the addition of data from additional detectors,
though the improvement can be small if the additional detectors are less sensitive than the original pair. In an effort to
improve convergence of the sampler, we have introduced dedicated proposals to update the right ascension and declination of the
source, $(\alpha, \delta)$, and the amplitude $A$, initial phase $\phi_0$ and orientation of the source, described by the polarization angle $\psi$ and
elipticity $\epsilon$. While the sky proposal we use is only strictly valid for a two detector network and for sources with
fixed elliptical polarization, it can be used for multi-detector networks, though the acceptance rate will be reduced in those cases.

The sky-ring proposal selects a new sky location $(\alpha_y, \delta_y)$ such that the time delay between the detectors are preserved. This requires an
overall time-shift $dt$ to be applied to the waveform, which can be computed from the difference in the time of arrival at each detector for the current  $(\alpha_x, \delta_x)$ and proposed  $(\alpha_y, \delta_y)$ location. The sky ring is defined by rotations about the vector $\hat{z}$ connecting the vertices of the two detectors.
Defining the angle between the current signal propagation vector $\hat{k}_x$ and $\hat{z}$:  $\theta={\rm acos}(\hat{z} \cdot \hat{k}_x)$, we can construct the orthonormal triad $(\hat{u}, \hat{v}, \hat{z})$ with
\begin{eqnarray}
\hat{u}  &= & \frac{\hat{k}_x - \cos\theta \hat{z}}{\sin\theta} \nonumber \\
\hat{v}  &= & \frac{\hat{z} \times \hat{k}_x}{\sin\theta}\, .
\end{eqnarray}
The new sky location is then found by drawing an angle $\phi$ uniformly in $[0,2\pi]$ and rotating about $\hat{z}$ to yield
\begin{equation}
\hat{k}_y = \sin\theta (\cos\phi \, \hat{u} + \sin\phi \, \hat{v}) + \cos\theta \hat{z} ,
\end{equation}
The proposal density for the sky ring move is constant, and cancels in the Metropolis-Hastings ratio. Since BayesWave
references the time of arrival to a reference detector (usually Hanford), there is no need to shift the arrival time. If,
however, the time of arrival is referenced to the Geocenter, then the arrival time needs to be shifted by an amount
$\Delta t = {\bf R}_i \cdot(\hat{k}_y  - \hat{k}_x)$, where ${\bf R}_i$ is the position of detector $i$ relative to
the Geocenter. If there are three or more detectors the sky ring proposal can be used by picking a pair of
detectors ${i,j}$. The acceptance of the sky ring proposal will be lower in multi-detector networks since the mapping
only keeps the arrival time constant for that one pair of detectors. 

The sky-ring proposal on its own has a low acceptance rate even for a two detector network. This is because the
projection of the signal onto the detectors depends on the sky location. For a two detector network, and for signals
with fixed elliptical polarization, we can find new extrinsic parameters, polarization angle $\psi$,
ellipticity $\epsilon$, amplitude $A$ and overall phase $\phi_0$ such that the waveforms in each detector are
identical at the new sky location. This mapping greatly improves the acceptance rate of the sky-ring proposal.

The requirement that the waveform projections are the same at the current and proposed sky location
yields the set of four equations:
\begin{eqnarray}
u_x f_{1+x}+v_x  f_{1\times x} &=& u_y f_{1+y}+v_y  f_{1\times y}  \nonumber \\
w_x f_{1+x}+z_x  f_{1\times x} &=& w_y f_{1+y}+z_y  f_{1\times y} \nonumber \\
u_x f_{2+x}+v_x  f_{2\times x} &=& u_y f_{2+y}+v_y  f_{2\times y} \nonumber \\
w_x f_{2+x}+z_x  f_{2\times x} &=& w_y f_{2+y}+z_y  f_{2\times y} 
\end{eqnarray}
where the $f_{i+}$ and $f_{i \times}$ are the primitive antenna patterns in the $i^{\rm th}$ detector, which are related to the full
antenna patterns by
\begin{eqnarray}
F_{+}(\alpha,\delta,\psi) &=& f_+(\alpha,\delta) \cos(2 \psi) + f_\times(\alpha,\delta) \sin(2 \psi)  \nonumber \\
F_{\times}(\alpha,\delta,\psi) &=&  - f_+(\alpha,\delta) \sin(2 \psi) + f_\times(\alpha,\delta) \cos(2 \psi) \, .
\end{eqnarray}
The quantities $u,v,w,z$ are defined such that
\begin{eqnarray}
u &=&  A(\cos\phi \cos(2\psi) +\epsilon \sin\phi \sin(2\psi))\nonumber \\
v &=& A(\cos\phi \sin(2\psi) -\epsilon \sin\phi \cos(2\psi)) \nonumber \\
w &=& A(\sin\phi \cos(2\psi) -\epsilon \cos\phi \sin(2\psi)) \nonumber \\
z &=&  A(\sin\phi \sin(2\psi) +\epsilon \cos\phi \cos(2\psi))
\end{eqnarray}
We can solve for $\{ u_y,v_y, w_y,z_y\}$:
\begin{eqnarray}
u_y &=& \frac{v_x {\rm f}[\times_y \times_x \times_x\times_y]  +    u_x {\rm f}[\times_y +_x  +_x \times_y]}{ {\rm f}[\times_y +_y +_y \times_y]    } \nonumber \\
v_y &=& \frac{v_x {\rm f}[\times_x +_y   +_y \times_x] +    u_x {\rm f}[+_x +_y +_y +_x] }{{ \rm f}[\times_y +_y +_y \times_y]  } \nonumber \\
w_y &=& \frac{z_x {\rm f}[\times_y \times_x \times_x\times_y]+    w_x{\rm f}[\times_y +_x  +_x \times_y] }{{  \rm f}[\times_y +_y +_y \times_y]  } \nonumber \\
z_y &=& \frac{z_x  {\rm f}[\times_x +_y   +_y \times_x] +    w_x {\rm f}[+_x +_y +_y +_x]}{{  \rm f}[\times_y +_y +_y \times_y] } 
\end{eqnarray}
where ${\rm f}[abcd] = f_{1 a} f_{2 b}  -f_{1 c} f_{2 d}$. Our next task is to invert the expressions for $(u,v,w,z)$ to solve
for $(A,\psi,\epsilon,\phi)$. Some algebra yields
\begin{equation}
\phi_y = \frac{1}{2} {\rm atan}\, (q)
\end{equation}
where
\begin{equation}
 q = \frac{2(u_y w_y + v_y z_y)}{(w_y^2+z_y^2)-(u_y^2+v_y^2)} \, .
\end{equation}
For $0 \leq \phi_y \leq  \pi/4$,  and $3\pi/4 \leq \phi_y \leq \pi$ the ellipticity is given by
\begin{eqnarray}
&&\epsilon_y^{-1} = \frac{u_y^2+v_y^2+w_y^2+z_y^2)\sqrt{1+q^2}+(u_y^2+v_y^2) }{2(u_y z_y - v_y w_y)\sqrt{1+q^2}}   \nonumber \\
&& \hspace*{0.5in} -\frac{(w_y^2+z_y^2)+2 q(u_yw_y +v_yz_y)}{2(u_y z_y - v_y w_y)\sqrt{1+q^2}} 
\end{eqnarray}
and the amplitude is given by
\begin{eqnarray}
&&A_y = \left( \frac{(\sqrt{1+q^2}+1)(u_y^2+v_y^2)+ 2 q(u_yw_y+v_y z_y)}{ 2 \sqrt{1+q^2}} \right. \nonumber \\
&&\hspace*{0.8in} \left.
+\frac{(\sqrt{1+q^2}-1)(w_y^2+z_y^2)}{ 2 \sqrt{1+q^2}} \right)^{1/2}
\end{eqnarray}
For $\pi/4 < \phi_y < 3\pi/4$ the ellipticity is given by
\begin{eqnarray}
&& \epsilon_y = \frac{(u_y^2+v_y^2+w_y^2+z_y^2)\sqrt{1+q^2}+(u_y^2+v_y^2)}{ 2(u_y z_y - v_y w_y)\sqrt{1+q^2}} \nonumber \\
&& \hspace*{0.5in}- \frac{(w_y^2+z_y^2)+2 q(u_yw_y +v_yz_y)}{ 2(u_y z_y - v_y w_y)\sqrt{1+q^2}} \, ,
\end{eqnarray}
and the amplitude is given by
\begin{eqnarray}
&&A_y = \left( \frac{(\sqrt{1+q^2}+1)(u_y^2+v_y^2)+ 2 q(u_yw_y+v_y z_y)}{ 2 \epsilon_y^2 \sqrt{1+q^2}} \right. \nonumber \\
&&\hspace*{0.8in} \left.
+\frac{(\sqrt{1+q^2}-1)(w_y^2+z_y^2)}{ 2  \epsilon_y^2 \sqrt{1+q^2}} \right)^{1/2}
\end{eqnarray}

Defining
\begin{eqnarray}
g &=& (1+\sqrt{1+q^2})v_y+q z_y \nonumber \\
h & =& (1+\sqrt{1+q^2})u_y+q w_y
\end{eqnarray}
the polarization angle is given by
\begin{equation}
\psi_y =
\left\{\begin{array}{lr}
 \frac{1}{2} \, {\rm atan_2}(g,h)     & \text{for }  0 \geq \phi_y \leq  \pi/4 \\
    & \\
 \frac{1}{2} \, {\rm atan_2}(\epsilon_y h, \epsilon_y g)        & \text{for }  \pi/4 < \phi_y \leq 3\pi/4 \\
 & \\
 \frac{1}{2} \,   {\rm atan_2}(-g,-h)       & \text{for }  3\pi/4 < \phi_y \leq \pi \\
  \end{array}\right.
\end{equation}

%\bence{Should we maybe move some of these nasty equations above to an appendix?}

Using the above formulae, the phase $\phi_y$ covers the range $[0,\pi]$. To cover the full range $\phi_y \in [0,2 \pi]$ we also need to include the solutions found by setting $\phi_y \rightarrow \phi_y+ \pi$ and
$\psi_y \rightarrow \psi_y+ \pi/2$, while keeping $A_y$ and $\epsilon_y$ fixed.

The deterministic mapping $\vec{x} \rightarrow \vec{y}$ given above requires a non-trivial Jacobian in the Metropolis-Hastings ratio given by $J=\vert \partial \vec{y} / \partial \vec{x}\vert$. Numerical central differences are then used to compute $\partial y^i /\partial x^j$

\begin{figure*}
\includegraphics[width=\textwidth]{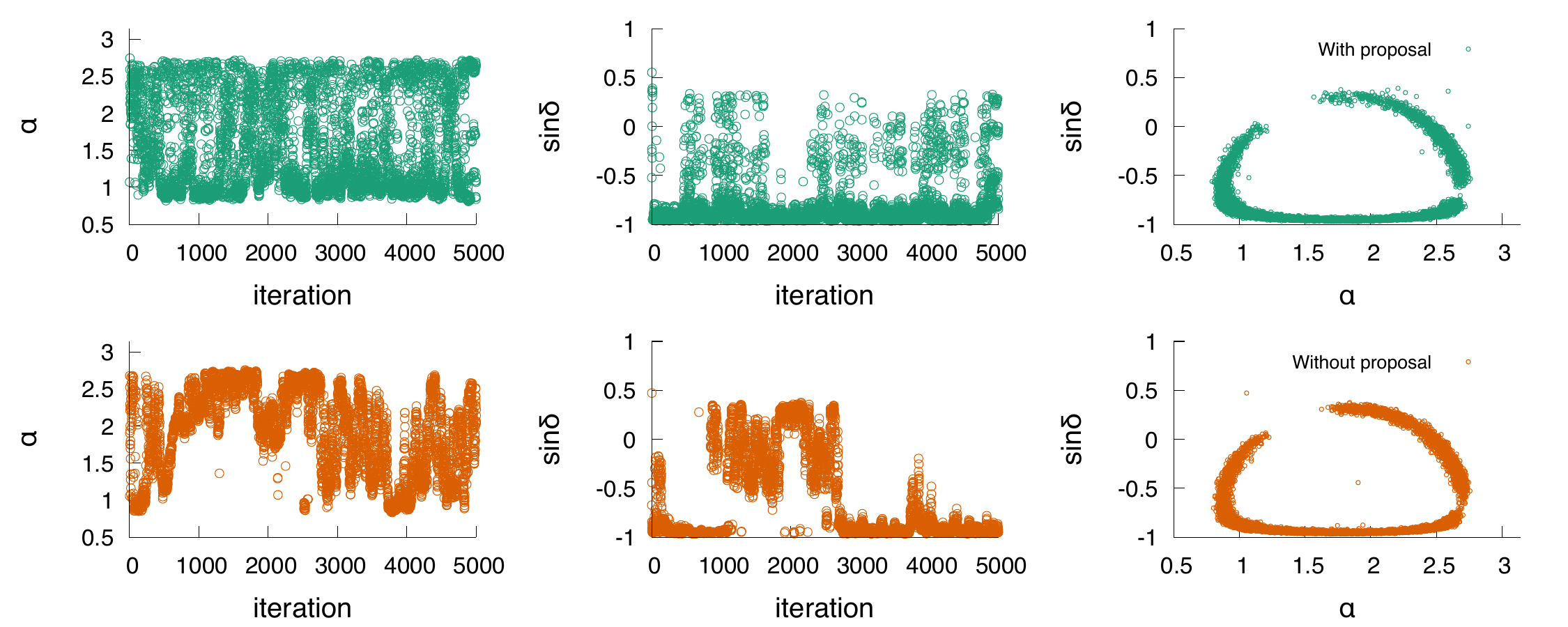}
\caption{Scatter plot of chain samples for sky localization parameters with (upper panel, green) and without (lower panel, orange) using the sky ring proposal. Shown here is a subset of a full chain to highlight the difference in mixing between the settings. The panels from left to right show the right ascension ($\alpha$) and sine of the declination ($\sin\delta$) as a function of chain iteration, and then the combined scatter plot of all chain samples.  We use the data containing GW150914 for this demonstration.}
\label{fig:sky-ring}
\end{figure*}

Figure \ref{fig:sky-ring} shows how the sky ring proposal improves the mixing of the Markov chain and therefore the convergence time of the analysis using GW150914 as an example.  The figure panels show the trace of the sky location parameters' ($\alpha$,$\sin\delta$) samples as the chain iterates. Without the sky ring proposal the sampler would have to run for longer to infer the correct relative weights between the different high-probability regions of the sky.

%%%%%%%%%%%%%%%%%%%%%%%%%%%%%%%%%%%%%%%%%%%%%%%%%%
\subsection{TFQ proposal}
\label{tfq}
%%%%%%%%%%%%%%%%%%%%%%%%%%%%%%%%%%%%%%%%%%%%%%%%%%

For trans-dimensional MCMC algorithms to efficiently sample in dimension space, well-designed proposal distributions which leverage domain knowledge are important. In addition to what is described in detail in \cite{Cornish:2014kda}, we have developed a new proposal to determine where wavelets should be placed in parameter space.

The proposal density is proportional to the matched filter signal to noise ratio $\SNR$ maximized over the wavelet phase, computed on a grid in time-frequency space with resolution of $5\ \rm{ms}$ in time and $4\ \rm{Hz}$ in frequency. The Q-scan is repeated for several different ``layers'' in $Q$, which has units of time, using a grid spacing of $2\ \rm{s}$. The result is a three-dimensional discretized grid in time-frequency-$Q$ space proportional to $\rho$. 
The distribution is then normalized by $\left(\sum_{ijk} \rho^2_{ijk}\right)^{1/2}$ where $i,j,$k are indices denoting the $t$, $f$, and $Q$ grid location.
The distribution is used to propose new wavelets in the fit, or to update current wavelet locations by rejection sampling uniform draws from $\{t,f,Q\}$ (TFQ) volume. The remaining wavelet parameters ($\mathcal{A}$ and $\phi_0$) are drawn from the prior.

\begin{figure}
\includegraphics[width=0.5\textwidth]{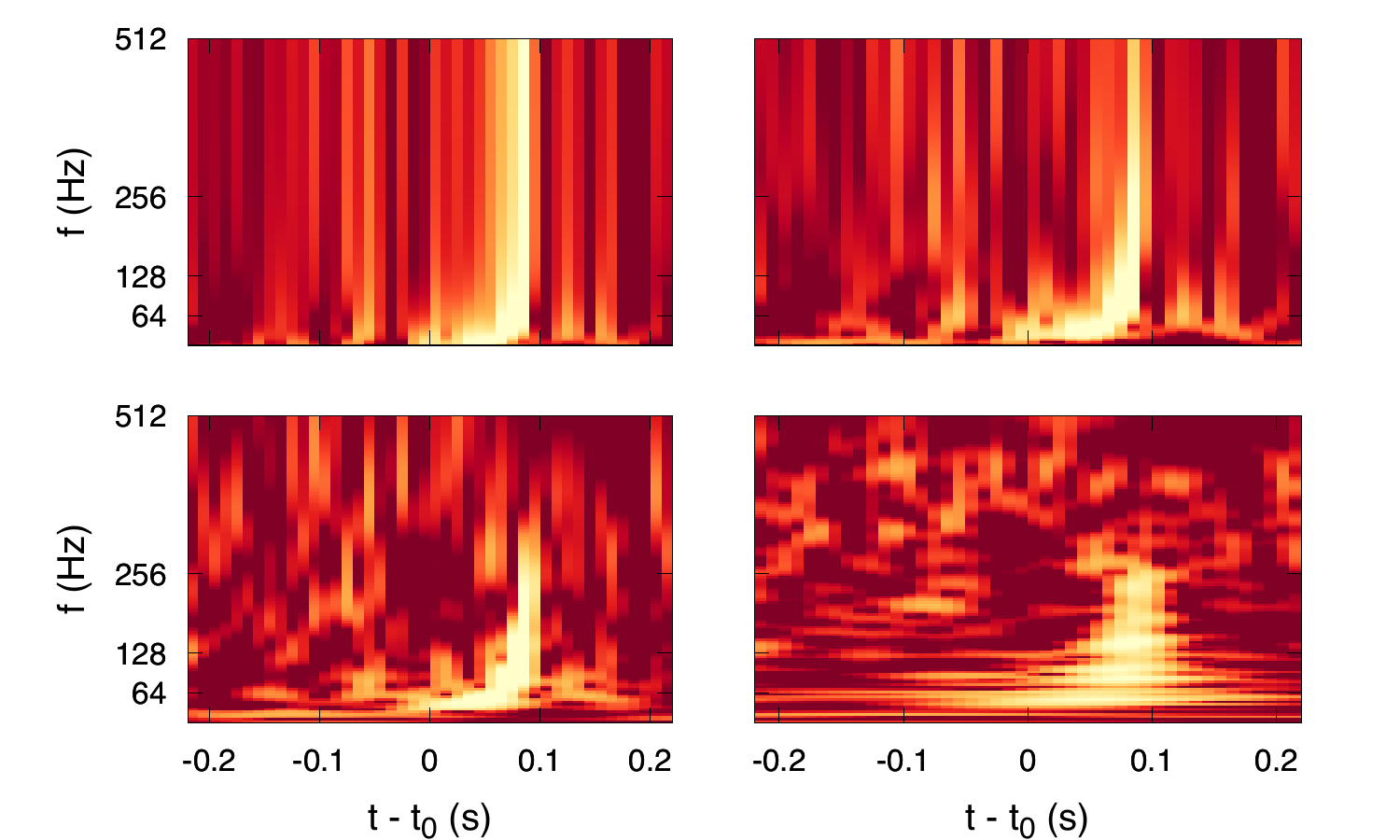}
\caption{Two dimensional slices in the time-frequency plane of the TFQ proposal at increasing $Q$ layers from top left to bottom right. The input data contains GW150914. The proposal will preferentially select draws from the $Q$ layers where the signal appears most compactly, thereby maximizing the $\SNR$ per cell in the grid.}
\label{fig:tfq}
\end{figure}

Figure~\ref{fig:tfq} shows two dimensional slices of the TFQ proposal at different $Q$ ``layers'' increasing from top left to bottom right.
Note how, as $Q$ changes, so too does the aspect ratio of features highlighted in the time-frequency map. 
Different signal morphologies will be better represented by different $Q$ layers, which in turn will provide better sampling efficiency. 
The example data used in this figure contains the BBH event GW150914.
The characteristic ``chirp'' shape is most clearly and compactly represented at the middle $Q$ layers shown here (top right and bottom left) and would therefore be preferentially chosen for proposing updates to the wavelet model. 

The TFQ proposal is most impactful as a proposal for transdimensional moves, when adding or removing a wavelet from the fit, but is also part of the proposal cycle for within dimension moves, taking an existing wavelet and proposing to replace it with a fair draw from the proposal.

%%%%%%%%%%%%%%%%%%%%%%%%%%%%%%%%%%%%%%%%%%%%%%%%%%
\subsection{Fast start PSD and glitch model}
\label{faststart}
%%%%%%%%%%%%%%%%%%%%%%%%%%%%%%%%%%%%%%%%%%%%%%%%%%

The  \BayesWave noise model may use many hundreds of parameters to describe the spline control points, the Lorentzian lines and the glitch model wavelets. Consequently, it can take hundreds of thousands of iterations for the sampler to reach the equilibrium distribution (``burn-in''), especially when loud glitches are present. The burn-in time can be reduced significantly by starting the chains at a good initial solution for the power spectral density and glitch model. We have adopted the fast deterministic method for iteratively estimating the PSD and finding a maximum likelihood solution for the glitch model that is part of the low latency \GlitchBuster algorithm, which can be used to remove noise transients from the LIGO/Virgo data in real time. 

The \GlitchBuster algorithm works as follows: The first step is to find a robust estimate for the power spectral density. Estimating the PSD for stationary, Gaussian noise is straightforward, and there are many methods to chose from. It is much more challenging to estimate the PSD for data that has a combination of Gaussian noise, non-stationary noise transients (glitches), and long duration non-stationarity that causes the PSD estimate to vary with time.  \GlitchBuster  employs an iterative approach using the same short data segments analyzed by \BayesWave. The first step is to apply a Tukey window then FFT the data to compute the power spectrum. A running median is used to smooth the spectrum. Choosing the width of the smoothing window involves a trade-off between spectral distortion and smoothness. If the window is too wide the smoothed spectrum underestimates the slope of the power spectrum, while if the window is too short the resulting spectrum will not be very smooth. We also want the window to be wide enough so that sharp spectral line features get flagged as outliers. For the short 4-8 second data segments that are typically analyzed by \BayesWave, we employ a 16 Hz window across much of the band, with a smaller 8 Hz window below 64 Hz, and a 4 Hz window below 32 Hz where the spectrum is very steep. With longer data segments the windows can be smaller since there are more frequency samples per Hz. Line features are identified as regions where the raw PSD exceeds the running median by some factor, typically set at 10. The full PSD model is the sum of the running median and the outliers. This initial PSD estimate can be biased by glitches, so the next step is to identify and remove the glitches. To do this the data is first whitened using the initial estimate for the PSD, then wavelet transformed using an over complete basis of continuous Morlet-Gabor wavelets. Wavelet denoising~\cite{TorrenceCompo} is then used to remove regions of excess power (the glitches). The denoised data is then returned to the frequency domain, and the PSD estimation procedure is repeated. The original data is whitened using the new PSD estimate, followed by another round of wavelet denoising. This cycle is repeated until the $\SNR$ of the glitch model stabilizes, which typically takes between one and five iterations. The entire procedure takes approximately one second for a four second data segment, taking proportionally longer as the segment length increases.

The PSD estimate, made up of a smooth component and a collection of outliers, is next mapped to the parameters used by the \BayesLine Bayesian spectral estimation algorithm. The smooth component of the PSD is used to compute the initial cubic spline model using a fixed frequency spacing, typically set equal to the minimum spacing allowed by the \BayesLine spline model. The outliers are mapped onto the Lorentzian line model by finding the central frequency, frequency extent and maximum height of each outlier region, and using these to compute the central frequency, amplitude and scale of the Lorentzian function that approximates the outliers.

\begin{figure}
\includegraphics[width=0.5\textwidth]{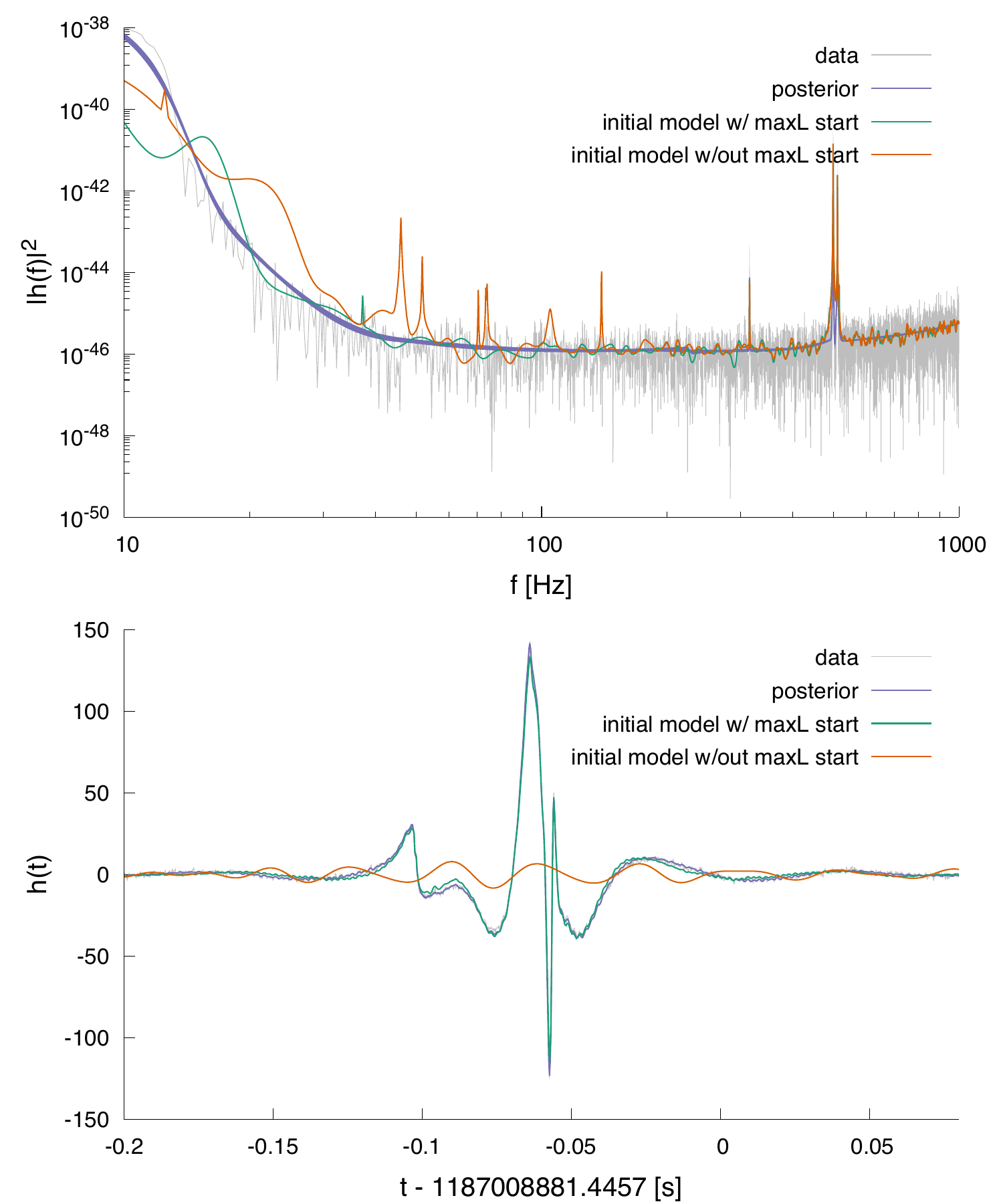}
\caption{Demonstration of the improved model initialization from \GlitchBuster using the glitch in the Livingston data near GW170817~\cite{TheLIGOScientific:2017qsa}. The top panel shows the data (gray), and PSD estimates with (green) and without (orange) the maximum likelihood wavelet initialization step from \GlitchBuster. These are compared to the posterior after the sampler has finished (purple).  The excess power in the PSD model shown in the orange curve is due to the \BayesLine parameters fitting part of the glitch.  The bottom panel shows the whitened glitch model compared to the data. The orange curve is from the initialization of the wavelet model that uses a fair draw from the prior. Note that the \GlitchBuster wavelet model starts at a solution where the reconstruction is consistent with the posterior. The median of the PSD posterior was used for whitening each of the glitch reconstructions in the bottom panel. }
\label{fig:glitchbuster}
\end{figure}

With the initial PSD model in hand, the next step is to solve for the glitch model in a form that can be used by \BayesWave. While the wavelet denoising procedure used in the PSD estimation produces a glitch model,  it is not in a form that can be used by \BayesWave. Moreover, the denoising is performed using wavelets with a single quality factor, and much better fits can be found using wavelets with a range of quality factors. To that end, the whitened data is wavelet transformed at geometric sequence quality factors using continuous Morlet-Gabor wavelets on a grid in time and frequency, following the identical procedure used to produce the TFQ proposal. The loudest pixel in the TFQ map is found, and if it exceeds some $\SNR$ threshold (e.g. $\SNR = 3$), the corresponding wavelet is subtracted from the data. Since continuous wavelets overlap with their neighbors, the TFQ map has to be updated in a region surrounding the wavelet that was removed. The procedure is repeated with the updated TFQ map until no significant outliers remain. The parameters of the loud wavelets identified in this way can now be used as a starting point for the \BayesWave glitch model. The iterative subtraction procedure is improved by adding a likelihood maximization step after each subtraction, similar to the F-statistic procedure~\cite{Jaranowski:1998qm}, but using the collection of wavelets as the filter functions. 

\begin{figure}
\includegraphics[width=0.5\textwidth]{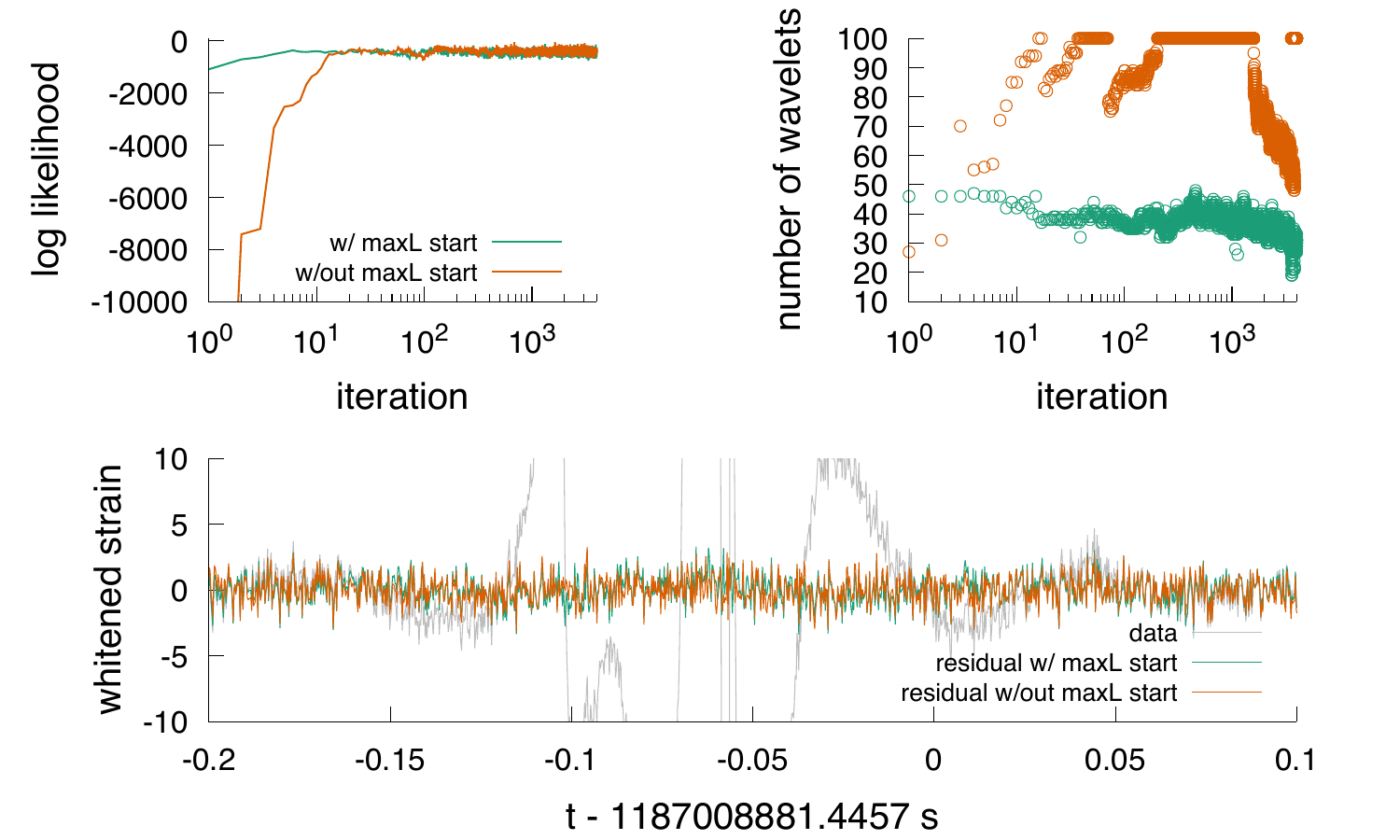}
\caption{Demonstration of convergence time improvement when using the PSD and glitch model initialization on a high $\SNR$ glitch in the Livingston detector during the binary neutron star merger GW170817.  The top left panel shows the log likelihood chain of the sampler, top right shows the number of wavelets used in the fit, and the bottom panel shows the residual after glitch subtraction.  The orange curves are for the version of \BayesWave using a random starting point for the chains.  The green traces are for the sampler that uses the new initialization step. Both versions achieve a similar fit to the glitch (as shown in the likelihood and residual plots) but the green curves achieve that fit with a smaller number of wavelets, which the orange curve was slowly trending towards before the sampler was stopped. }
\label{fig:freqwave}
\end{figure}

Figure~\ref{fig:freqwave} compares the convergence on data containing a high amplitude noise transient with (green) and without (orange) using the \GlitchBuster initialization. The top left panel shows the likelihood chain for the sampler, both of which reach similar values although the \GlitchBuster-initialized chain achieves that value $\sim\mathcal{O}(10)$ times faster.  However, the number of wavelets used by the naive start (top right panel) is larger, and 10s of thousands of sampler iterations were required for the model to even begin sampling with the more parsimonious number of wavelets.  The bottom panel shows that in both cases the glitch is adequately removed from the data (gray) by comparing the residuals (orange and green), also indicated by the comparable likelihoods achieved by both chains. The data used in this example contains the glitch just before the GW170817 merger~\cite{TheLIGOScientific:2017qsa}.

%%%%%%%%%%%%%%%%%%%%%%%%%%%%%%%%%%%%%%%%%%%%%%%%%%
\section{Post processing}
\label{sec:post_processing}
While the output of the RJMCMC in \BayesWave is samples from the posterior distribution of wavelet parameters and extrinsic parameters common to the signal (such as sky location, polarisation information), ultimately it is the morphology and properties of the GW signal that are of interest. Below we discuss the post processing used to translate raw samples into more meaningful outputs.

\subsection{Waveform reconstructions}
For each sample in the resulting posterior, the wavelet parameters are summed to produce a GW waveform $\bold{h}=h(t_i)$, where $i \in \left\{1,N_t\right\}$ with $N_t$ the number of discrete time steps\footnote{The number of discrete time steps is determined by the sampling rate and segment length used in the \BayesWave analysis}.  For each time step $t_i$, the \BayesWave post processing calculates the median and bounds of the $50\%$ and $90\%$ credible intervals on the posterior distribution of $h(t_i)$.  An example of a waveform reconstructions is shown in Fig.~\ref{fig:tfTracks}.

\BayesWave also produces a posterior distribution on $\tilde{h}(f)$, the GW waveform in the frequency domain.  The median and credible intervals for $\tilde{h}(f)$ are calculated in the same manner as described above.

\subsection{Frequency evolution reconstructions}\label{sec:ft}
\begin{figure}[]
    \includegraphics[width=\hsize]{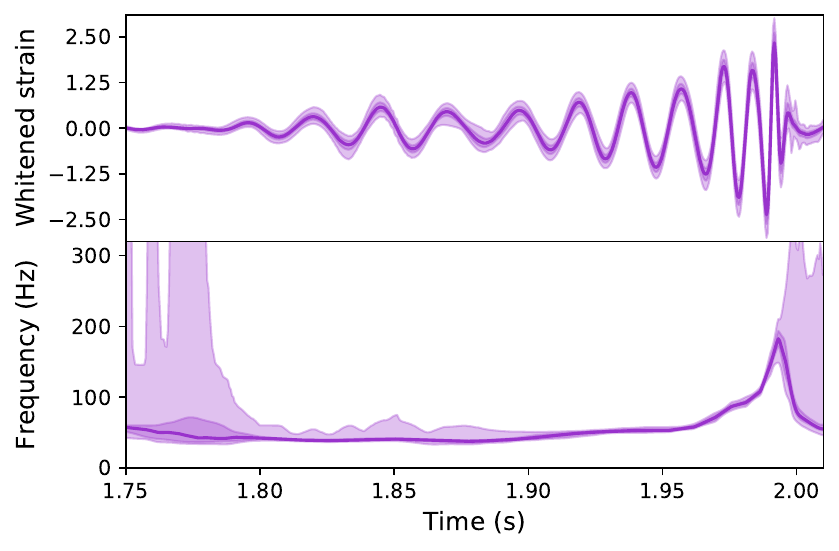}
        \caption{Example of the reconstructions from the post processing phase for the whitened time domain waveform (top), and the frequency evolution over time (bottom) for the example event GW170814~\cite{Abbott:2017oio}.  The solid line is the median reconstruction, and the shaded bands are the $50\%$ and $90\%$ credible intervals.  For the frequency evolution over time, the process described in Sec.~\ref{sec:ft} uses the zero-crossings of the reconstructed waveform, and as such times where the reconstructed strain is consistent with zero the $f(t)$ reconstruction will have large credible intervals in this region.  This is seen in times before about 1.8 seconds, and after merger (about 2.0 seconds) in the lower plot. 
          }
    \label{fig:tfTracks}
\end{figure}

In addition to the time and frequency domain waveforms, one may also be interested in looking at the frequency evolution over time of the GW signal (for example, one may wish to look for the characteristic ``chirp'' of a compact binary inspiral).  The \BayesWave post processing produces posterior distributions of this frequency evolution, $f(t)$.  This $f(t)$ is found by using points where $h(t) = 0$ (also called the zero crossings).  The frequency at the $j^\mathrm{th}$ zero crossing, denoted $t^0_j$ is given by
\begin{equation}
  f(t^0_j) = \frac{1}{t^0_{j+1}-t^0_{j-1}}.
\end{equation}
The \BayesWave post processing calculates $f(t)$ for each $h(t)$ posterior sample, and again calculates the median and $50\%$ and $90\%$ credible intervals.  An example is shown in Fig.~\ref{fig:tfTracks}.

\subsection{Waveform moments}

Another way to characterize the signal is to calculate the central moments of the reconstructed waveform. {\tt BayesWave} post processing calculates the first two central moments both in the time domain (central time and duration) and the frequency domain (central frequency and bandwidth) along with their probability distributions. In principle, we can calculate higher order moments too, but those are less intuitive, and expected to be measured less precisely. A comprehensive study of the parameter estimation capabilities of {\tt BayesWave} (including estimation of waveform moments) has been presented in \cite{becsy:2017}.

\subsection{Whitening tests}

\begin{figure}
\includegraphics[width=0.5\textwidth]{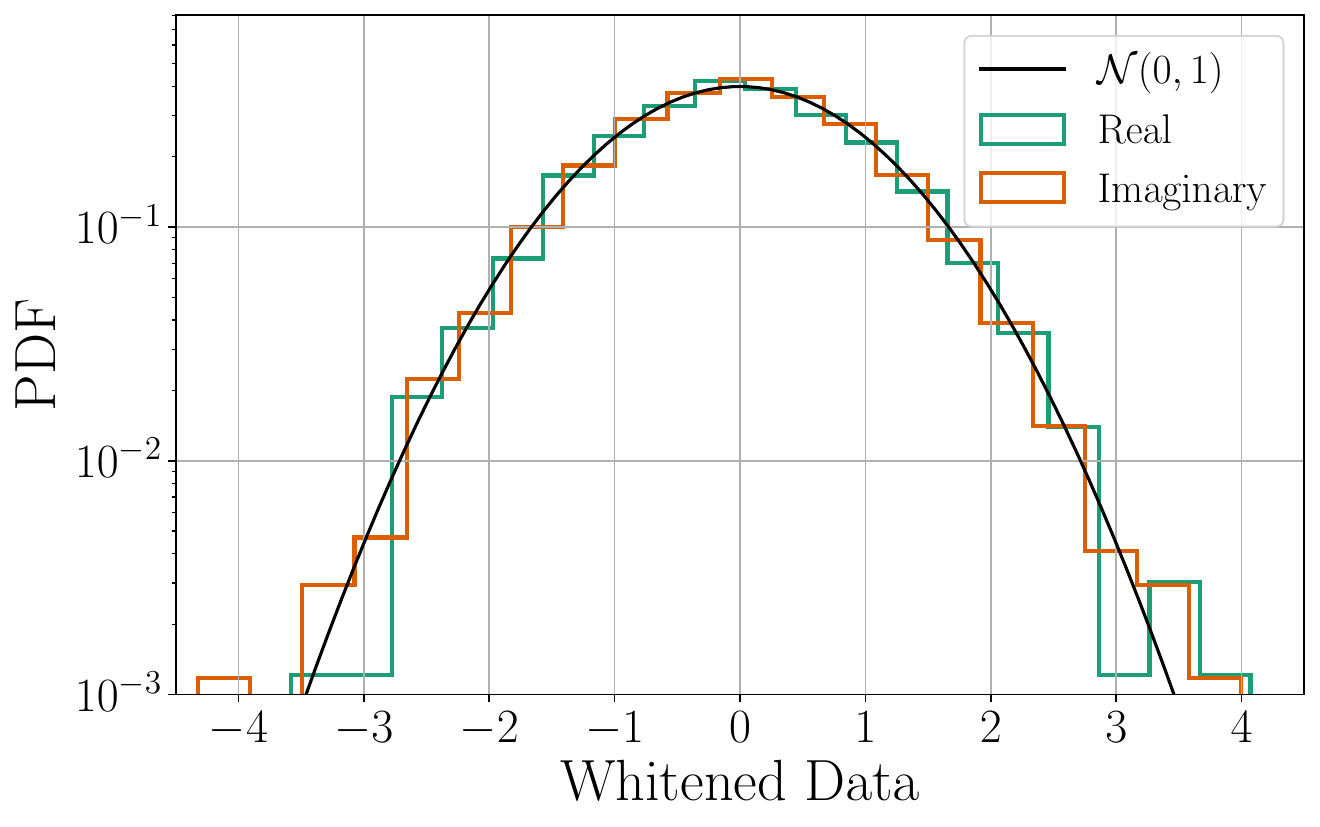}
\caption{Histogram of the real and the imaginary part of the
whitened Fourier residuals of 4s of data around GW150914. The noise power spectral density
has been computed with \BayesWave and the target ${\cal{N}}(0,1)$ distribution is shown for reference.
\label{residuals}}
\end{figure}

When the PSD is estimated from on-source data using the {\tt BayesLine} algorithm, a number of tests are performed in order to ensure
that the computed PSD whitens the data. The tests are described in more detail in~\cite{Chatziioannou:2019} and include histograms of the 
real and imaginary Fourier domain residuals as compared to a ${\cal{N}}(0,1)$ distribution, properties of the combined 2-D distribution, and the Anderson-Darling test. The latter quantifies the degrees to which samples are drawn from a target distribution and results in a p-value for the null hypothesis that the 
Fourier residuals are draws from ${\cal{N}}(0,1)$. The Anderson-Darling test is performed for various bandwidths, the ensure that the resulting PSD 
whitens all relevant frequency ranges of the analysis. An example of a histogram of the whitened Fourier residuals is shown in
Fig.~\ref{residuals} corresponding to 4s of LIGO-Hanford data around GW150914. The p-value for the plotted data is $0.81$, suggesting
that the null hypothesis that the residuals are drawn from ${\cal{N}}(0,1)$ cannot be ruled out.

%%%%%%%%%%%%%%%%%%%%%%%%%%%%%%%%%%%%%%%%%%%%%%%%%%

%%%%%%%%%%%%%%%%%%%%%%%%%%%%%%%%%%%%%%%%%%%%%%%%%%
\section{Review tests}
\label{sec:review}
%%%%%%%%%%%%%%%%%%%%%%%%%%%%%%%%%%%%%%%%%%%%%%%%%%

\BayesWave and \BayesLine periodically undergo standard review tests for sampling algorithms. Perhaps the most common
(and computationally inexpensive) review test is the ``constant likelihood test", i.e. run the code in a configuration where the likelihood function
is a constant. In this case the posterior is equal prior and the resulting samples must follow the prior distributions for all model parameters.
This test ensures that sampling satisfies detailed balance and it does indeed produce fair samples from the posterior distribution. We routinely 
confirm that \BayesWave passes this test.

Another common test (though computationally more expensive) that checks the likelihood itself is the ``P-P test", where simulated signals are
drawn from the prior distribution of each parameter and then analyzed. If a sampler is unbiased, then the true value for each parameter must be 
at the $p-th$ percentile of its posterior distribution for $p$ events. Stated differently, a plot of the fraction of events where the true value is at a certain 
percentile of the posterior must be diagonal for each parameter. This test is stronger but more computationally intensive, requiring dozens of injections.

\begin{figure}
\includegraphics[width=0.5\textwidth]{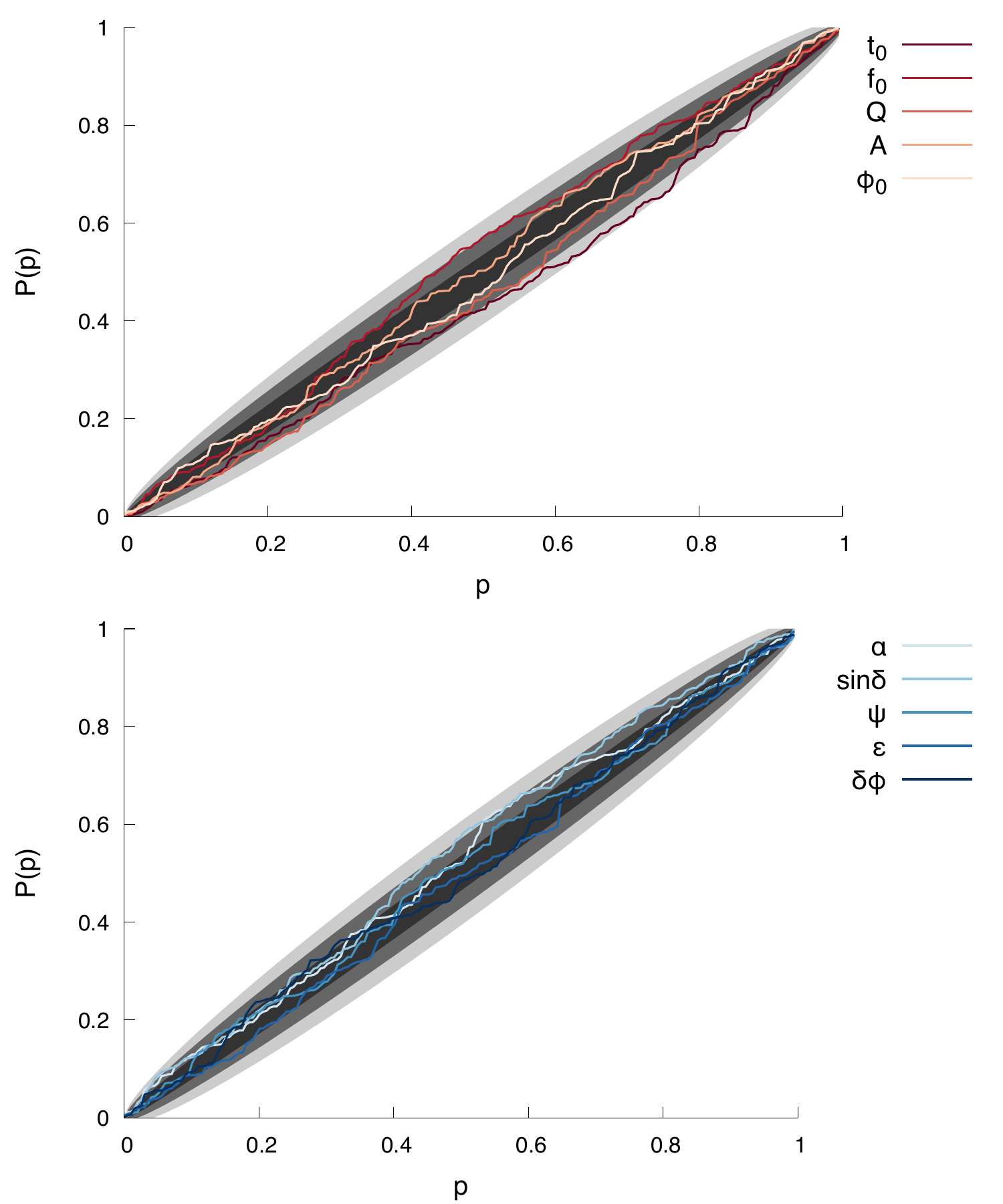}
\caption{P-P plots for model parameters and sky localization.}
\label{pp-plot}
\end{figure}

Figure~\ref{pp-plot} shows the results of the P-P test for the sky location (top panel) and the individual model parameters (wavelet parameters and 
extrinsic parameters) (bottom panel). Shaded regions denote 1-, 2-, 3-$\sigma$ errors. In all cases we recover diagonal lines within the expected error. The study was performed using simulated signals and simulated noise.

%%%%%%%%%%%%%%%%%%%%%%%%%%%%%%%%%%%%%%%%%%%%%%%%%%
\section{Conclusions}
\label{sec:conclusions}
%%%%%%%%%%%%%%%%%%%%%%%%%%%%%%%%%%%%%%%%%%%%%%%%%%

The \BayesWave algorithm continues to be improved, with additional functionality added and efficiency achieved since the initial release.  The most significant changes in the release described here include: the ability to model signals with general polarization content; simultaneous modeling of signals and noise transients; and significantly improved sampling. With this added functionality the use cases for  \BayesWave continue to grow, including glitch subtraction, tests of general relativity, and independent checks of the waveform models used to describe compact binary mergers.

Work continues to further extend and improve the performance. Future updates will allow for the joint sampling of compact binary coalescence templates and noise transients in addition to marginalization over the power spectral density of the noise. This functionality will be particularly valuable for low mass systems, such as binary neutron star mergers, where the long duration of the signals makes it highly likely that noise transients will also be present in the data. New tests of general relativity will be made possible by extending the polarization model to allow for scalar and vector polarization states. Dynamic spectral modeling, where the power spectral density can change with time, will be made possible by switching the analysis from the Fourier domain to the discrete wavelet domain~\cite{Cornish:2020odn}.  

Looking further afield, the \BayesWave approach is being applied to other branches of gravitational wave astronomy, including pulsar timing arrays~\cite{Ellis:2016mtg,Becsy:2020utk} and the future space-based LISA detector~\cite{Robson:2018jly}.

%%%%%%%%%%%%%%%%%%%%%%%%%%%%%%%%%%%%%%%%%%%%%%%%%%%
\appendix

%%%%%%%%%%%%%%%%%%%%%%%--------------------------------------------------------------------------
\section{Software \& Workflow}
\label{sec:AppA}

\BayesWave is an open source software project distributed under the terms and conditions of the Gnu Public License (GPL2).  Source code and documentation can be found at~\cite{bayeswave_src}.  Code development takes place via GitLab flow, with the core development team contributing and enhancing the software through regular merge requests from feature branches on their own forks of the repository.

The \BayesWave software stack and workflow is comprised of 4 principal components: \bayeswavepipe (written in python): a workflow generation tool; \BayesWave (C): the main analysis application which reads and conditions data, and fits for the various models described in this work; \BayesWavePost (C): post-processing application which parses the results from \BayesWave to produce reconstructed time- and frequency-domain representations of the models and cleaned data.  Results are presented through a collection of python-based plotting and web-page generation scripts.  Given a single time to analyze, along with a configuration file detailing the time-frequency volume of interest and the analysis mode, \bayeswavepipe is used to construct a simple HTCondor\cite{htcondor} DAGMan workflow in which each of these applications is executed in serial.  The process is trivially extended to larger analyses, such as Monte-Carlo simulations signal detection and characterization or background trials for significance estimation, by executing the same workflow as many times as required but as parallel, independent jobs on high-throughput computing resources.

\BayesWave is packaged with conda~\cite{bayeswave_feedstock} and technically- and scientifically-reviewed releases are available from the standard IGWN conda environment~\cite{igwn_conda} which is distributed via the OASIS CVMFS~\cite{cvmfs} repository managed by the Open Science Grid project~\cite{osg07,osg09}.  Docker images with the most recent release and the current state of the main git branch are available from the container registry associated with the source code repository.  Finally, these docker images are distributed to the OSG CVMFS container repository (where they are automatically converted to singularity images), provisioning access to the latest and development versions of the code, in addition to the reviewed packages in conda, on OSG resources.

%%%%%%%%%%%%%%%%%%%%%%%--------------------------------------------------------------------------
\section{Workflow Characterization}
\label{sec:AppB}

In this appendix we characterize typical \BayesWave and \BayesLine analysis job
profiles to provide an idea of the computational resources required.  Exemplary
scenarios selected include power spectral density estimation for BBH-like
events using \BayesLine and gravitational wave signal model-only analysis, both
in a variety of time-frequency configurations used for previous LIGO-Virgo
detections.  While there is some further variation when evaluating the glitch
or signal-plus-glitch models, the signal-only model serves, to first order, to
illustrate the resource consumption per model when running \BayesWave.  For
example, the wall-time to compute a signal-versus-glitch Bayes factor for one
of the configurations shown here can be estimated from the \BayesLine wall
time, plus $2\times$ the \BayesWave wall time.

Our figures of merit for each configuration are the cumulative wall clock- and
CPU-times, the peak memory usage and the peak disk usage, recorded by HTCondor
over the lifetime of the job.  Workflows used in this characterization ran
from Singularity containers, using native support in HTCondor, and images
deployed in the OSG CVMFS repository.  All calculations were performed on Intel
Xeon E3-1240 v5 processors in the dedicated LIGO computing cluster at the
California Institute of Technology.

Finally, it is worth noting that, due to its long run-times (up to nearly 2
days in some cases considered here and often longer in more extreme cases),
\BayesWave saves its the state of the calculation every hour and exits, to be
resumed by the workflow management system, in order avoid data loss in case
of worker node contention or loss of connectivity.  We find this process has
negligible detrimental impact on the time-to-solution for continuously running
jobs, while being absolutely critical to operating on shared resources.

\subsection{BBH PSD Estimation}
\BayesLine power spectral density estimates have proven a critical component of
upstream, template-based parameter estimation efforts.  To characterize
\BayesLine performance we have re-analyzed 24 BBH events from the GWTC-2
catalog~\cite{Abbott:2020niy} using identical time-frequency configurations to
those used in the LIGO-Virgo parameter estimation studies\footnote{The O3a
catalog events were analyzed with a variety of time-frequency configurations:
higher-mass BBH mergers result in shorter-duration, lower-frequency signals,
necessitating analysis of a smaller time-frequency volume.}.  Each event was
analyzed with ten independent trials to account for fluctuations in system
load, and network / disk performance.  Two sets of results are produced: one
set using the ``low-latency'' \BayesLine configuration, with 100000 MCMC
iterations, which was used to inform online parameter estimation efforts in O3,
and a ``high-latency'' \BayesLine configuraion with 4000000 MCMC iterations.

\begin{figure}[]
    \includegraphics[width=\hsize]{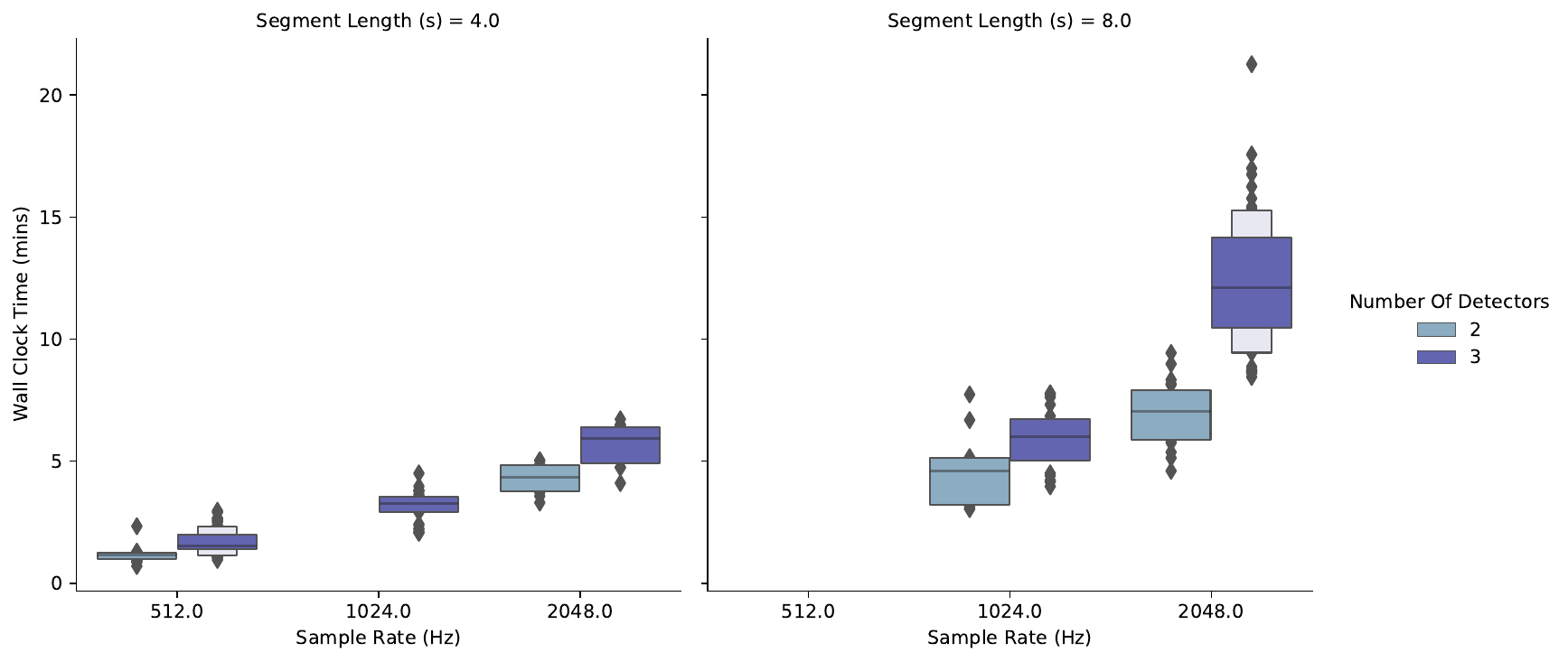}\\
    \includegraphics[width=\hsize]{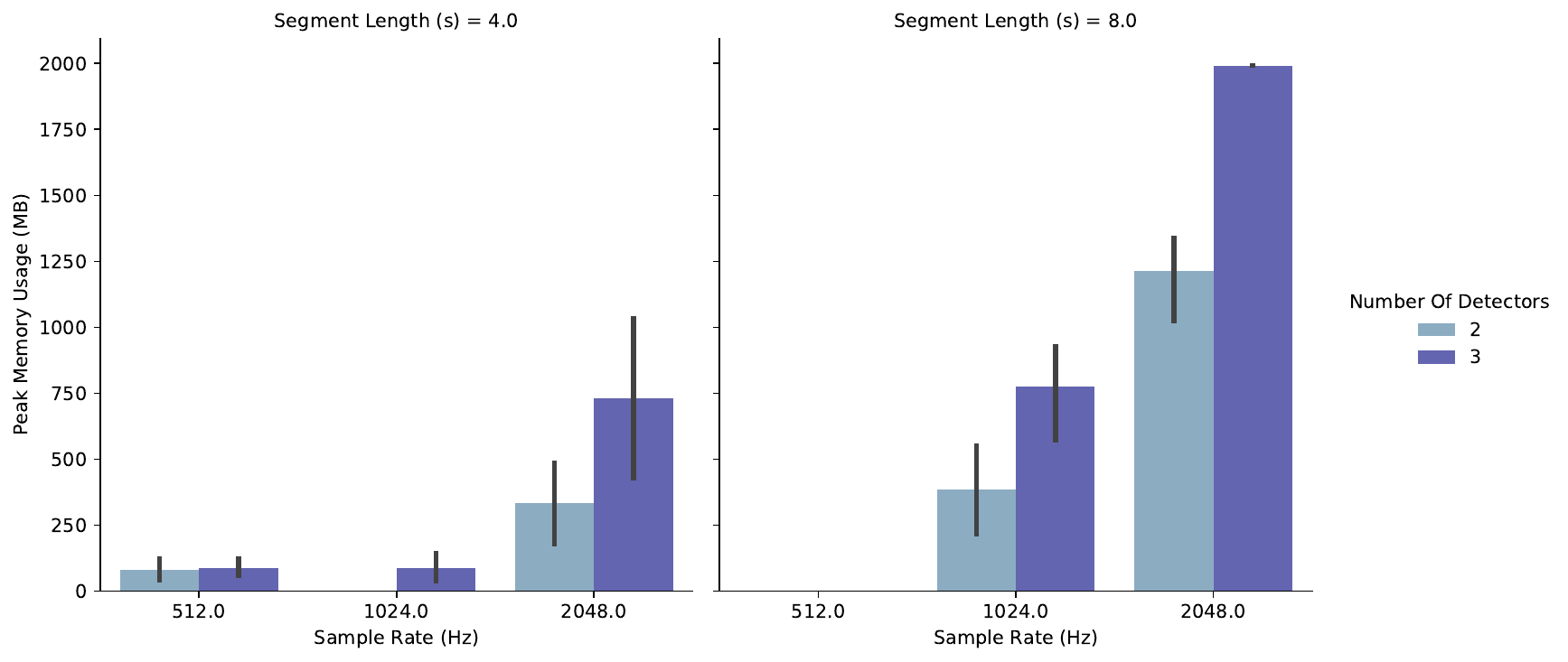}\\
    \includegraphics[width=\hsize]{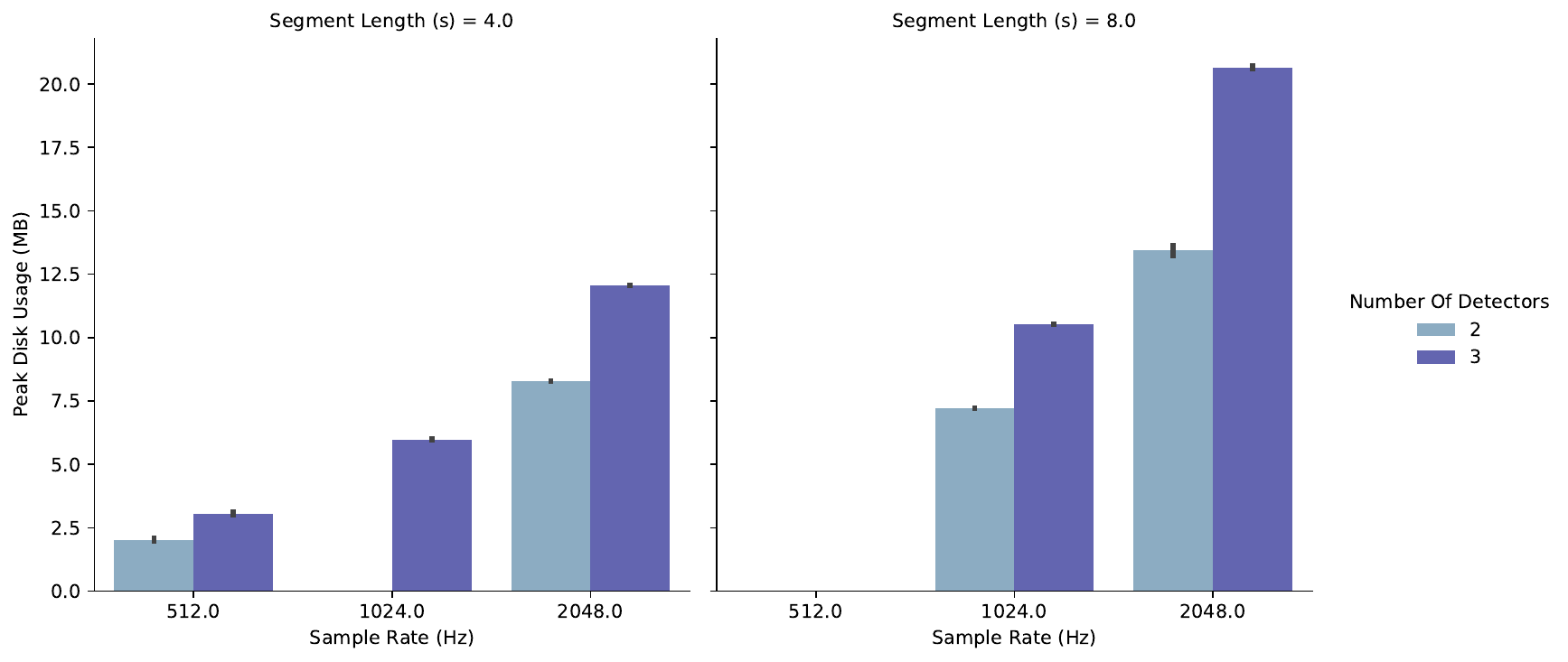}\\
    \caption{Job characterization metrics for the main \BayesWave RJMCMC process for
    low-latency \BayesLine PSD estimation for LIGO-Virgo O3a catalog BBH events.
    Results are grouped by each event's time-frequency and detector configuration.}
    \label{fig:psdFastProfileBayesWave}
\end{figure}

Figure~\ref{fig:psdFastProfileBayesWave} show, from top to bottom: the
wall-time, peak memory usage and peak disk space used by the main \BayesWave
RJMCMC program, running in the low-latency \BayesLine configuration.  CPU-time
is not measurably different from wall time for these jobs and is not shown
here.  Results from each of the 24 BBH events are grouped by the \BayesLine
time-frequency configuration and number of detector data streams used in the
analysis.   The median wall-time for low-latency \BayesLine PSD estimation
ranges between $\sim 1$\,minute, for the smallest time-frequency volumes
considered here (2 data streams with duration 4s and sample frequency
512\,Hz), to $\sim 12$\, minutes for the 3 detector analyses with 8 second
segments, sampled at 2048\,Hz.  Note that the corresponding \BayesWavePost
jobs, which parse the sampled model parameters and reconstruct the inferred
PSD, typically complete in under 30\,seconds here, with minimal memory and disk
footprints; their data is not shown.

\begin{figure}[]
    \includegraphics[width=\hsize]{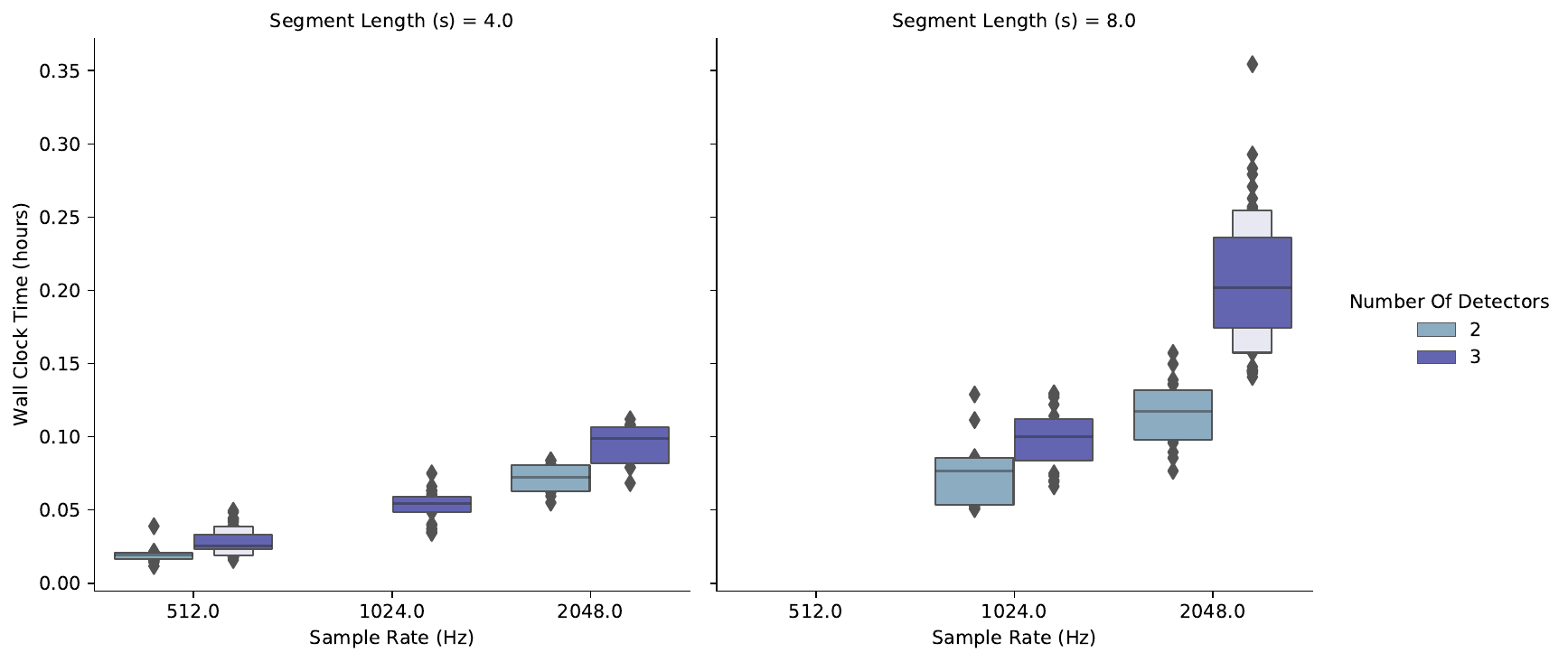}\\
    \includegraphics[width=\hsize]{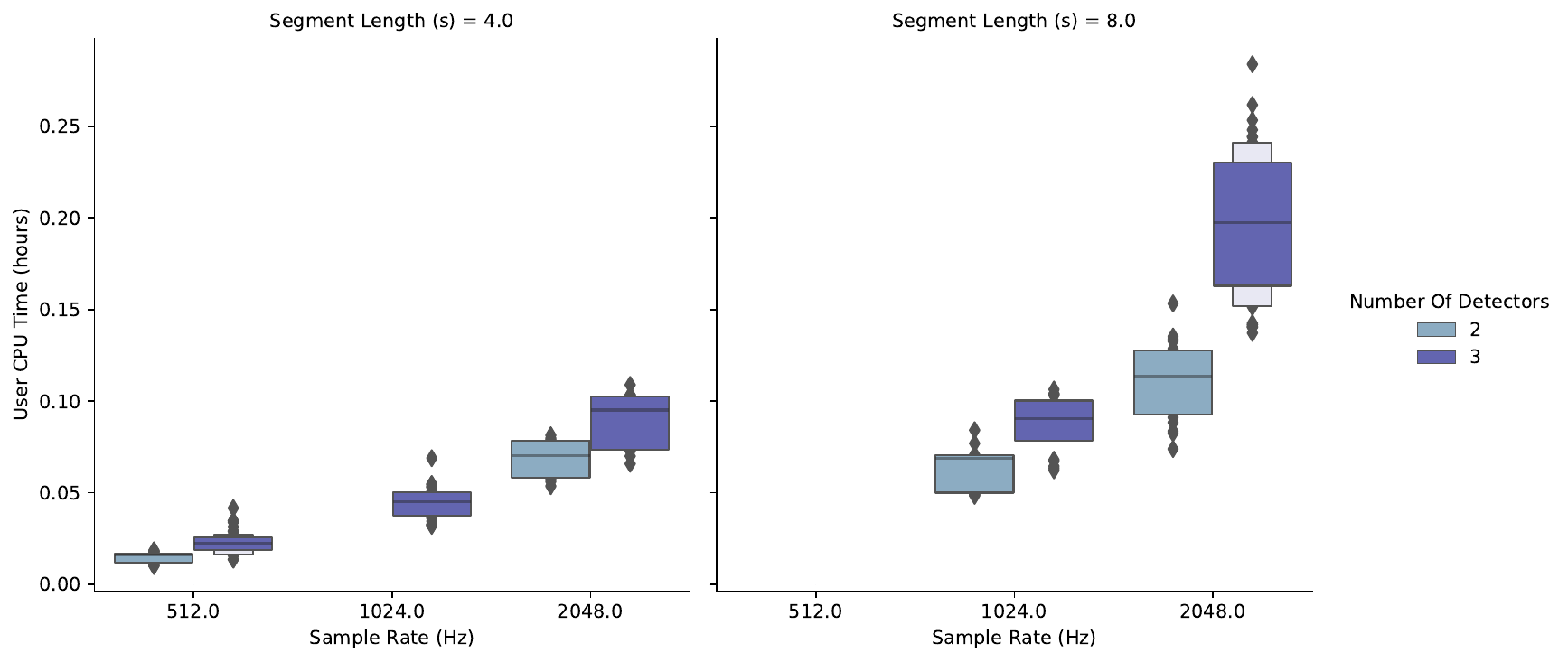}\\
    \includegraphics[width=\hsize]{BayesWave_ResidentSetSize_RAW_MB_configs.pdf}\\
    \includegraphics[width=\hsize]{BayesWave_DiskUsage_RAW_MB_configs.pdf}\\
    \caption{Analysis job metrics for the main \BayesWave RJMCMC process for final \BayesLine PSD estimation for LIGO-Virgo O3a catalog BBH events.}
    \label{fig:psdSlowProfileBayesWave}
\end{figure}

Equivalent statistics for the offline \BayesLine PSD estimation are shown in
figure~\ref{fig:psdSlowProfileBayesWave}.  The time-to-solution has increased
proportionately with the 40-fold increase in the number of MCMC iterations,
with wall times now measured in hours: the PSD for the smallest time-frequency
volume is now only attained after a median wall time of 40\,minutes, while the
longer 8-second segments require 7.5\,hours.  

Now that we are into a regime where jobs are running long enough as to require
periodic checkpointing their progress, it is worth checking that the
exit/resume behavior every hour is not detrimental to workflow efficiency.
Figure~\ref{fig:psdSlowProfileBayesWave} also shows the CPU-time, as well as
the wall time:  The difference in the median Wall- and CPU-time is less than
5\% of the total Wall-time, while their distributions are, to all intents and
purposes, identical. This provides some reassurance that there is no
significant startup penalty incurred from saving and resuming jobs.  

The difference in the memory footprint is more marginal, with that cost being
dominated by storing a identical time-frequency maps of the data.
Unsurprisingly, the storage requirements have also increased significantly,
going from $\sim$2--20\,MB for the low-latency analysis, to $\sim$18--130\,MB
for the offline analysis.

As before, the \BayesWavePost jobs do not add significantly to the total
time-to-solution, taking, at most, abou5 15\,minutes to complete and presenting
nearly identical memory and disk footprints as the parent \BayesWave jobs.

\subsection{BBH Waveform Reconstructions}
We now repeat the characterization above using the gravitational wave signal
model.  In these analyses, we compute the evidence for the signal model, sample
the posterior probability distribution function for the signal model parameters
to reconstruct the underlying gravitational wave signal present in the data
stream from each detector, and we compute posterior probability density
functions for a variety of moments of the reconstructed waveforms.  Typically,
the waveform reconstructions from such analyses are used to search the data for
deviations from physically parameterized waveform models~\cite{}.  As before,
we re-analyze each of the BBH events from the LIGO-Virgo O3a catalog~\cite{},
using 10 independent trials in each case to average over any fluctuations in
e.g., system load or network performance.  These waveform reconstruction
analyses use the median offline \BayesLine PSDs computed in the previous
section.  That is, the metrics obtained here are purely for the evaluation of
the signal model and do not include any overhead for PSD estimation.

Finally, it should be noted that full \BayesWave waveform reconstruction
comparisons, like those in \S~VIII of~\cite{Abbott:2020niy}, typically use
Monte-Carlo simulations of template-based waveform reconstructions, in addition
to analyzing the actual gravitational wave signal.  In assessing any expected
total resource consumption based on the following measurements then, the reader
should scale these results by the number of Monte-Carlo simulations to be
performed (usually $\mathcal{O}(100)$).

\begin{figure}[]
    \includegraphics[width=\hsize]{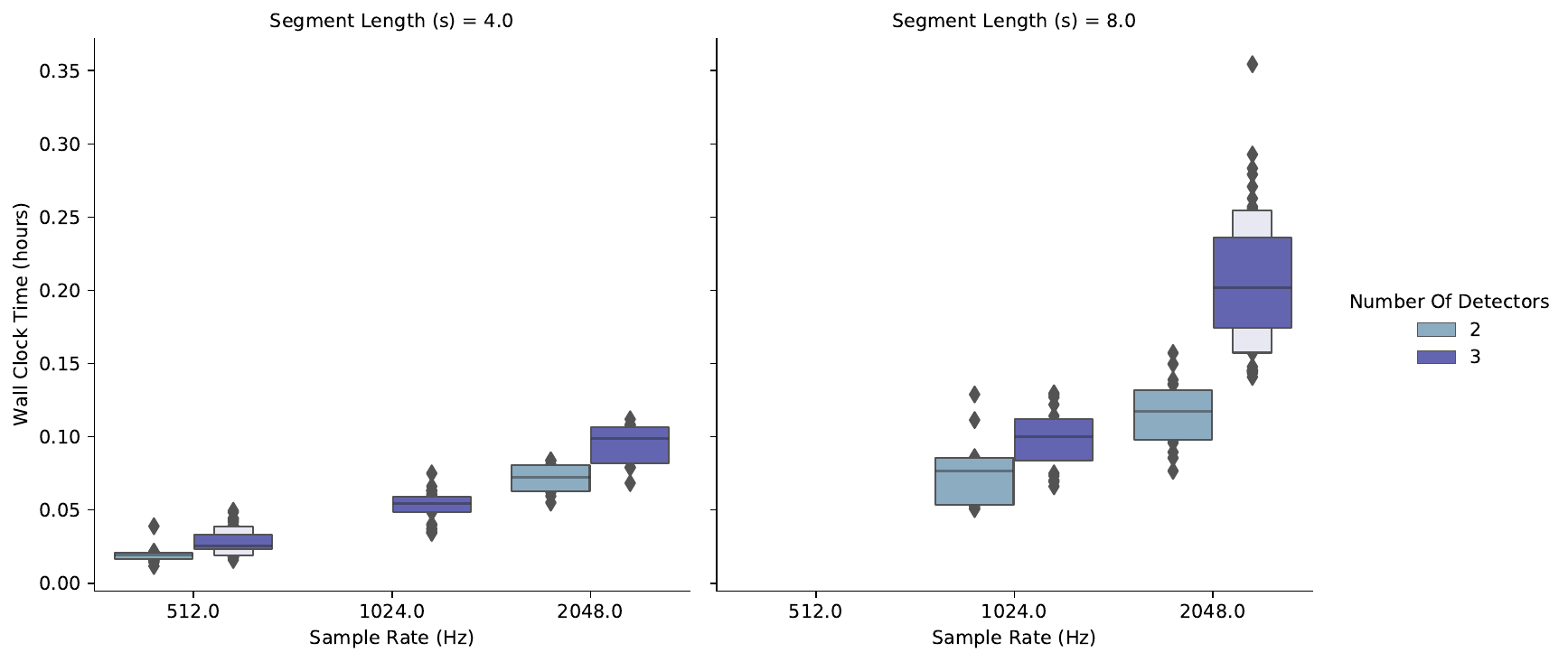}\\
    \includegraphics[width=\hsize]{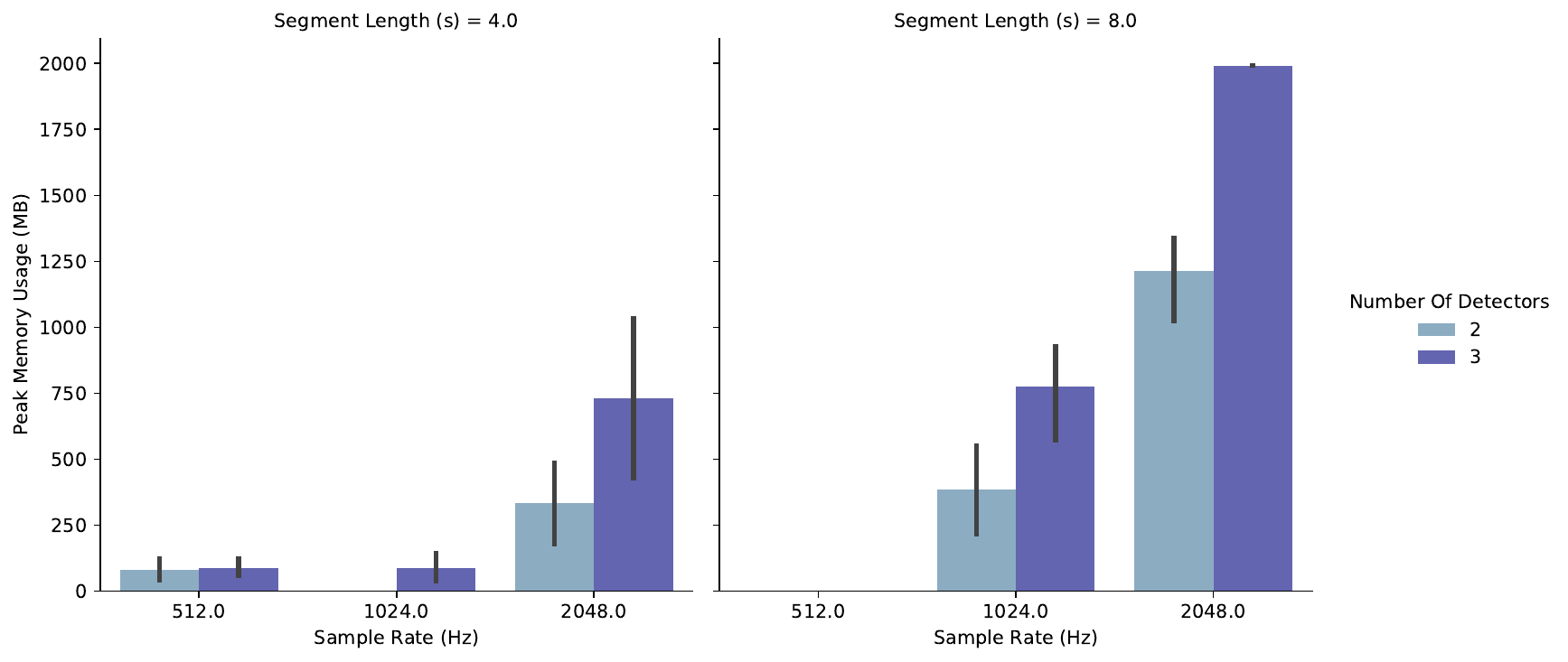}\\
    \includegraphics[width=\hsize]{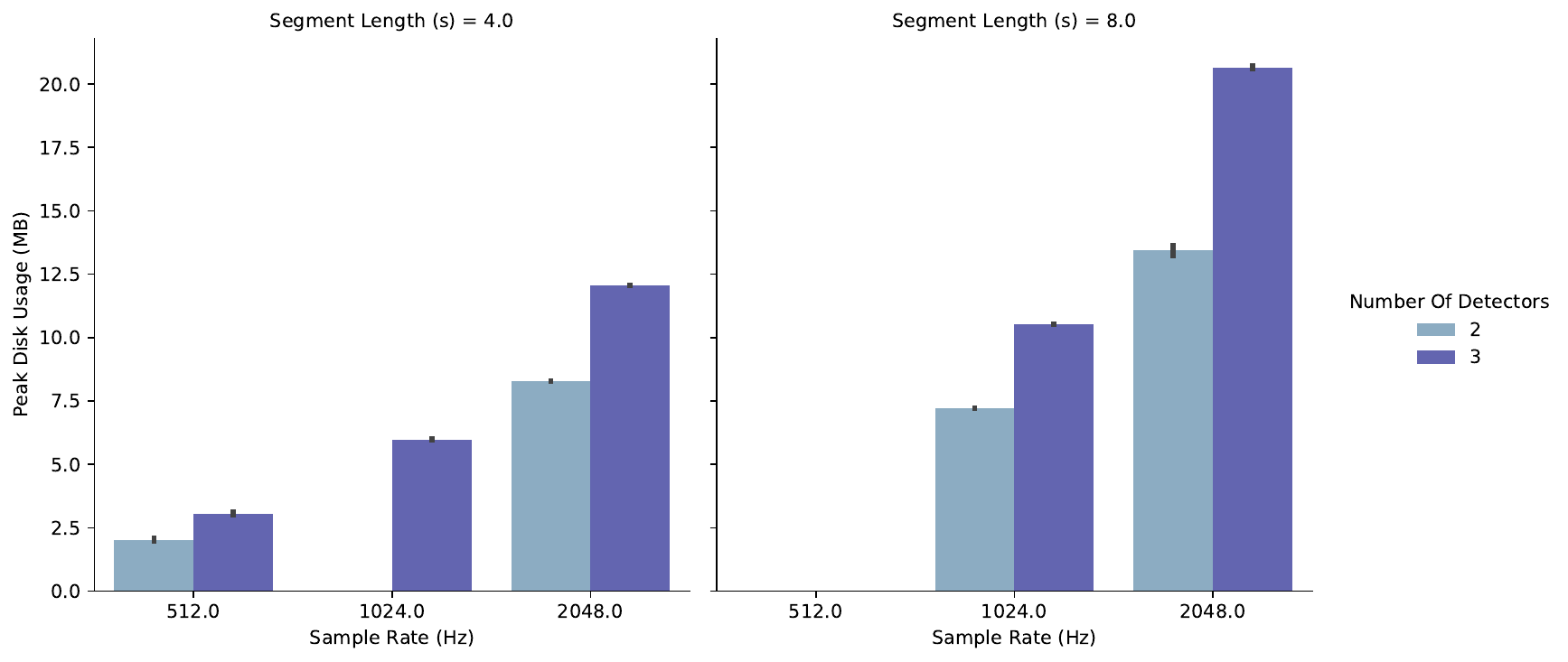}\\
    \caption{Analysis job metrics for the main \BayesWave RJMCMC process for evaluation of the gravitational wave signal model for the purposes of waveform reconstruction analyses.  Configurations shown are those used for the LIGO-Virgo O3a catalog events.}
    \label{fig:sigRecProfileBayesWave}
\end{figure}

Figures~\ref{fig:sigRecProfileBayesWave} show the wall-time, peak memory
consumption and the disk usage of the BBH events from the O3a catalog.  Again,
results are grouped by total time-frequency volume used in the analyses.  As
before, the difference in wall- and cpu-times, is much smaller than the overall
wall-time so, in the interests of brevity, the latter is not shown.  Evaluating
the signal model now presents a more formidable computational challenge, with
wall-times in the range of 5--48\,hours.  Further,
figure~\ref{fig:sigRecProfileBayesWave} shows the metrics for the corresponding
\BayesWavePost child jobs.  While these jobs still typically complete in well
under an hour, their memory requirements can be substantial, due to a need to
store several thousand time-domain waveform reconstructions in memory - the
median peak memory consumption gets as high as 11\,GB for the 8\,second
segments considered here.

\begin{figure}[]
    \includegraphics[width=\hsize]{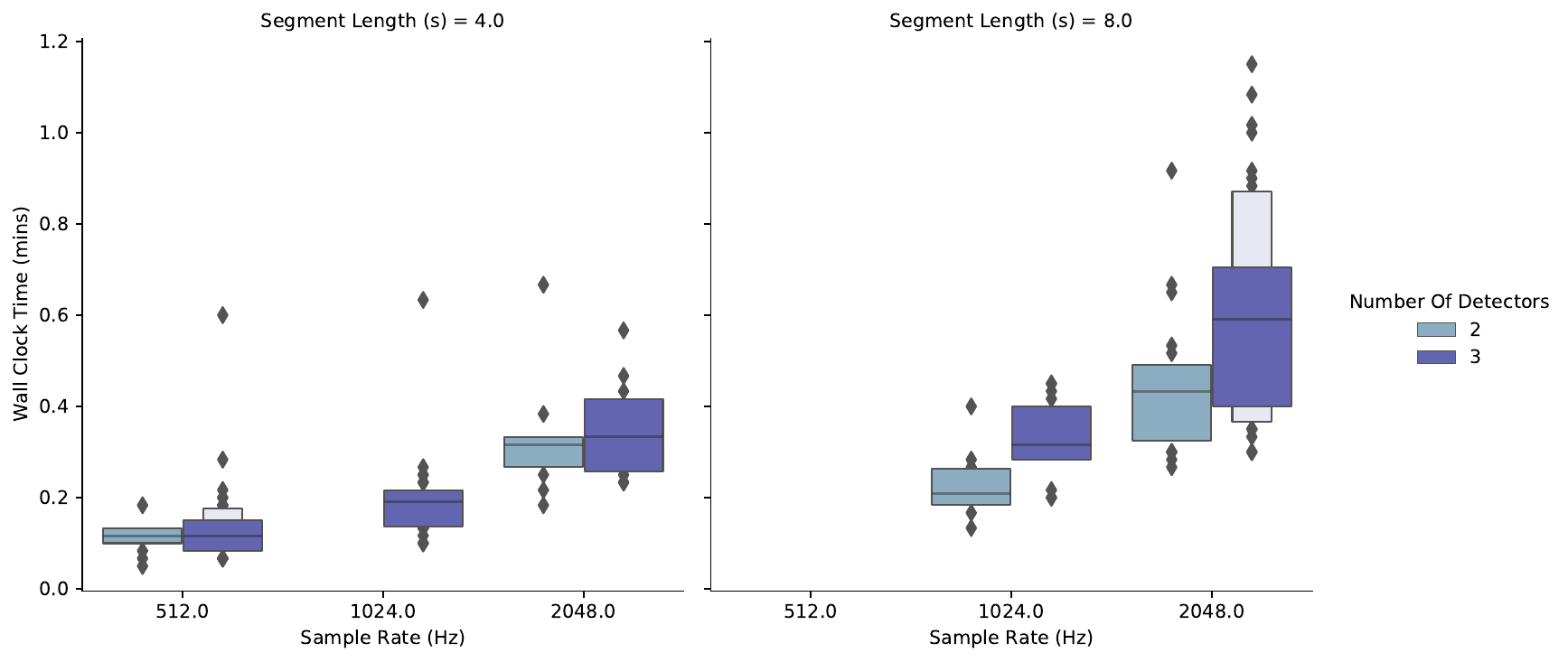}\\
    \includegraphics[width=\hsize]{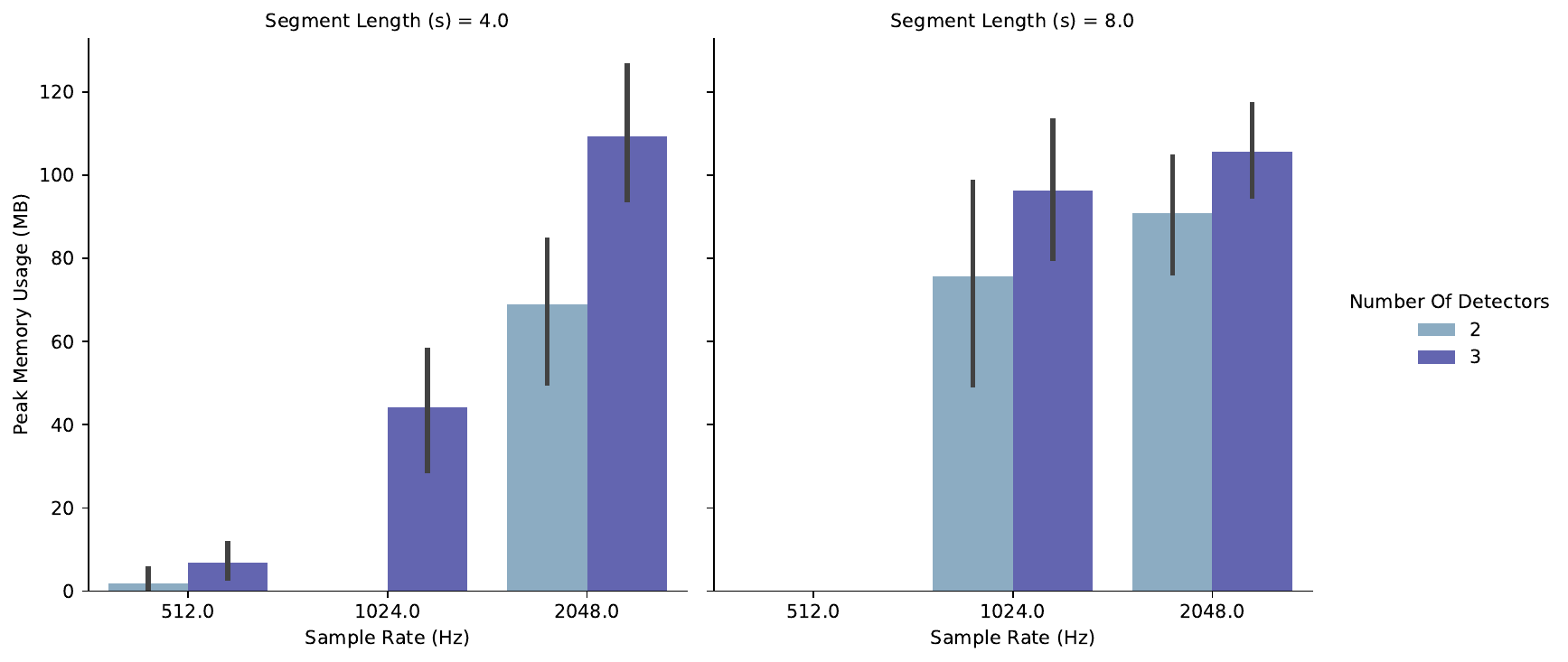}\\
    \includegraphics[width=\hsize]{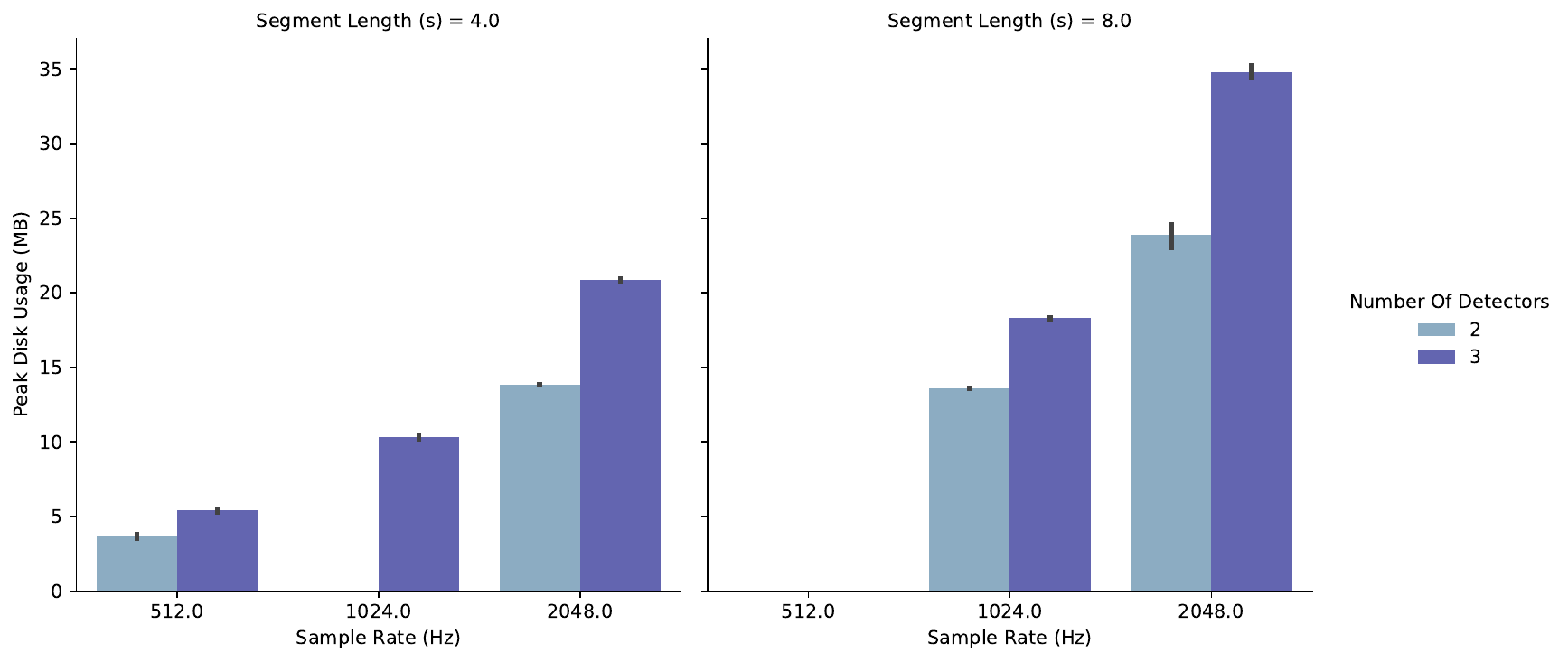}\\
    \caption{Analysis job metrics for the \BayesWavePost process, run subsequent to the jobs shown in figure~\ref{fig:sigRecProfileBayesWave}.}
    \label{fig:sigRecProfileBayesWavePost}
\end{figure}

%%%%%%%%%%%%%%%%%%%%%%%--------------------------------------------------------------------------
\section{Optimization}
\label{sec:AppC}

\subsection{Delta Likelihood Updates}

Computing the full likelihood (\ref{eq:like}) for a network with $M$ detectors over a time span with $N$ data points takes ${\cal O}(MN)$ operations. The computational cost of the likelihood calculation can be significantly reduced by performing ``delta'' updates that are localized in frequency. For example, the amplitude envelope of the Morlet-Gabor wavelets falls off as $\exp[-\pi^2\tau^2 (f-f_0)^2]$, and to a good approximation, the likelihood only needs to be computed for frequencies $ f_0 -\delta f < f  <  f_0 +\delta f $, with $\delta f \simeq 4/(\pi \tau) = 8 f_0/Q$ when updating the contribution from a particular wavelet. 

The delta likelihood updates work as follows. Suppose the model is currently given by $h_x$. The residual is $r_x = d - h_x$ and the log likelihood is given by
$\ln L_x = -(r_x|r_x)/2 + W_x$ where $W_x$ depends only on the noise model. Holding the noise model fixed and adding (removing) a single wavelet $\Psi$ to the model yields the updated residual $r_y = r_x \mp \Psi$ and the updated likelihood
\begin{equation}
\ln L_y = \ln L_x + \delta \ln L
\end{equation}
where
\begin{equation}
 \delta \ln L = (\Psi|\Psi)/2 \pm (\Psi | r_x) \, .
\end{equation}
The delta likelihood only has to be evaluated over frequency band $f_0 \pm \delta f$. The computational saving is significant for wavelets with large quality factors $Q$, and/or for wavelets with low central frequencies $f_0$. The same delta likelihood method can be used to update the parameters of an existing wavelet in the model by first subtracting a wavelet with the current parameters, then adding a wavelet with the updated parameters.

\subsection{Recursive Evaluation of Wavelets}

The expressions (\ref{mg}) for the Morlet-Gabor wavelets contain real and complex exponentials that can be costly to evaluate. Replacing these function calls by recursion relations significantly decreases the computational cost. The term $\exp[\pm 2\pi i (f-f_0)t_0 +\phi_0)]$ in the frequency domain expression, and $\cos(2\pi f_0 (t-t_0) +\phi_0)$ in the time domain expression, can be computed using trigonometric recursion relations. Picking a reference frequency $f_*$ or reference time $t_*$, the phase terms can be written as $\Phi_n = \Phi_* + n \Delta \Phi$, where for the frequency domain case $\Phi_* = 2\pi (f_*-f_0)t_0 + \phi_0$ and $\Delta \Phi = 2\pi \Delta f t_0$, and for the time domain case  $\Phi_* = 2\pi (t_*-t_0)f_0 + \phi_0$ and $\Delta \Phi = 2\pi \Delta t f_0$. The evaluation is initialized by computing the four terms $\cos\Phi_0$, $\sin\Phi_0$,  $\cos\Delta \Phi$, $\sin\Delta \Phi$. Subsequent values are found by multiplication and addition according to the recursion relation
\begin{eqnarray}
\cos \Phi_{n+1}= \cos \Phi_{n} \cos\Delta\Phi - \sin\Phi_{n}  \sin\Delta\Phi \nonumber \\
\sin \Phi_{n+1}  = \cos\Phi_{n} \sin\Delta\Phi + \sin\Phi_{n} \cos\Delta\Phi  \, .
\end{eqnarray}

The amplitude terms contain exponentials of the form $\exp[-Q^2 f/f_0]$, $\exp[-\pi^2\tau^2 (f-f_0)^2]$, and $\exp[-(t-t_0)^2/\tau]$. The first of these terms is linear in the frequency increment $\Delta f$, while the later include quadratic terms in $\Delta f$ or $\Delta t$. The term $A = \exp[-Q^2 f/f_0]$ can be computed using the recursion
\begin{equation}
A_{n+1} = A_n e^{-Q^2 \Delta f/f_0}\, ,
\end{equation}
with $A_0= \exp[-Q^2 f_*/f_0]$.
The terms with quadratic in the increments require a two-part recursion. For example, the frequency domain term $A= \exp[-\pi^2\tau^2 (f-f_0)^2]$ can be computed using the recursion relation
\begin{eqnarray}
A_{n+1} &=& A_n \alpha_n e^{(2 f_0 \Delta f - \Delta f^2)\pi^2\tau^2} \nonumber \\
\alpha_{n+1} &=& \alpha_n e^{-2(\pi \tau \Delta f)^2}
\end{eqnarray}
with
\begin{eqnarray}
A_0 &\equiv& e^{-(\pi^2\tau^2(f_*-f_0)^2)} \nonumber \\
\alpha_0 &=& e^{-2 \pi^2 \tau^2 f_* \Delta f} \nonumber \\
\end{eqnarray}
The quadratic  time-domain amplitude $\exp[-(t-t_0)^2/\tau]$ can be computed in a similar fashion.

 %%%%%%%%%%%%%%%%%%%%%%%%%%%%%%%%%%%%%
\acknowledgements

This research has made use of data, software and/or web tools obtained from the Gravitational Wave Open Science Center (https://www.gw-openscience.org), a service of LIGO Laboratory, the LIGO Scientific Collaboration and the Virgo Collaboration.
LIGO is funded by the U.S. National Science Foundation.
Virgo is funded by the French Centre National de Recherche Scientifique (CNRS), the Italian Istituto Nazionale della Fisica Nucleare (INFN) and the Dutch Nikhef, with contributions by Polish and Hungarian institutes.
NJC and BB acknowledge the support of NSF awards PHY1607343 and PHY1912053.
The authors are grateful for computational resources provided by the LIGO Laboratory and supported by National Science Foundation Grants PHY-0757058 and PHY-0823459, and for resources provided by the Open Science Grid~\cite{pordes:2007,Sfiligoi:2009}, which
is supported by the National Science Foundation award 1148698, and the U.S.
Department of Energy's Office of Science.
The Flatiron Institute is supported by the Simons Foundation.
The GT authors gratefully acknowledge the NSF for
financial support from awards PHY 1806580, PHY
1809572, and TG-PHY120016.
Parts of this research were conducted by the Australian Research Council Centre of Excellence for Gravitational Wave Discovery (OzGrav), through project number CE170100004.
\bibliography{OurRefs}

\end{document}